\documentclass[a4paper,10pt]{article}

\usepackage{amssymb, amsthm}
\usepackage{amsmath}
\usepackage[usenames,dvipsnames]{color}
\usepackage{float, graphicx}
\usepackage[colorlinks, citecolor=blue, linkcolor=blue, urlcolor=Maroon, filecolor=Maroon]{hyperref}
\usepackage{epsfig,psfrag}
\usepackage[bf]{caption}
\usepackage{multirow}
\usepackage{booktabs}
\usepackage[abs]{overpic}

\usepackage{geometry}
\allowdisplaybreaks[4]
\newcommand{\sfrac}[2]{{\textstyle\frac{#1}{#2}}}
\newcommand{\und}{\qquad\textrm{and}\qquad}
\newcommand{\RR}{{\mathbb{R}}}

\newcommand{\dd}{{\rm d}}

\titlepage
\title{
\vspace*{-1.6cm}
\begin{flushright}
\normalsize{ITP--UH--05/12}\\
\end{flushright}
\vskip2cm
 Heterotic string plus five-brane systems with asymptotic AdS$_3$
\vspace{1cm}}
\author{Karl-Philip Gemmer${}^\dagger$,\ Alexander S. Haupt${}^\dagger$,\  Olaf~Lechtenfeld${}^{\dagger\times}$,\\ Christoph
N\"olle${}^\dagger$ \ and \
Alexander~D.~Popov${}^{*}$}
\date{}

\begin{document}

\setlength{\baselineskip}{12pt plus 0.3pt minus 0.1pt} 
\maketitle
\thispagestyle{empty}
\begin{center}
\noindent ${}^\dagger${\em
Institut f\"ur Theoretische Physik,
Leibniz Universit\"at Hannover \\
Appelstra\ss{}e 2, 30167 Hannover, Germany }\\
{Emails: Karl-Philip.Gemmer, Alexander.Haupt, Olaf.Lechtenfeld, Christoph.Noelle@itp.uni-hannover.de}
\\[6mm]
${}^\times${\em
Centre for Quantum Engineering and Space-Time Research \\
Leibniz Universit\"at Hannover \\
Welfengarten 1, 30167 Hannover, Germany }
\\[6mm]
\noindent ${}^*${\em
Bogoliubov Laboratory of Theoretical Physics, JINR\\
141980 Dubna, Moscow Region, Russia}\\
{Email: popov@theor.jinr.ru}
\end{center}

\vspace{2mm}
\vspace{1.5cm}

\begin{abstract}
 We present NS1+NS5-brane solutions of heterotic supergravity on curved geometries.
They interpolate between a near horizon AdS$_3\times X^k\times
\mathbb T^{7-k}$ region and $\mathbb R^{1,1}\times c(X^k)\times \mathbb T^{7-k}$, where $X^k$ (with $k =3,5,6,7$) is a
$k$-dimensional geometric Killing spinor manifold, $c(X^k)$ its Ricci-flat cone and $\mathbb T^{7-k}$ a $(7-k)$-torus.
The solutions require first order $\alpha'$-corrections to the field equations, and special point-like instantons play
an important role, whose singular support is a calibrated submanifold wrapped by the NS5-brane.
It is also possible to add a gauge anti-5-brane. We determine the super isometries of the near horizon geometry which
are supposed to appear as symmetries of the holographically dual two-dimensional conformal field theory. 
\end{abstract}
\numberwithin{equation}{section}

\parindent=0cm
\parskip=\bigskipamount
\newpage
\setcounter{page}{1}

\section{Introduction}

 Brane solutions of 10- and 11-dimensional supergravities have played an important role in the development of string
theory since the second superstring revolution, when it was realized that besides 1-dimensional extended objects, string
theory also requires the inclusion of higher-dimensional branes. The near horizon geometry of a supergravity $p$-brane
usually consists of a $(p+2)$-dimensional anti-de Sitter space times a compact manifold, and the AdS/CFT
correspondence relates the supergravity solution to a $p$-dimensional conformal field theory on the conformal boundary
of the anti-de Sitter space, which is supposed to govern the dynamics of a decoupled brane in some gravitational
background. 
The most prominent branes are listed in Table \ref{tab:oldbranes}:
 \renewcommand{\arraystretch}{1.3}
  \begin{table}[H]\centering
 \begin{tabular}{c|c|c|c}  \toprule
  &  SUGRA & near horizon geometry & distant vacuum \\ \hline 
 M5-brane & 11D & AdS$_7\times X^4$ &  $\mathbb R^{5,1} \times c(X^4)$  \\
 M2-brane & 11D &  AdS$_4\times X^7$ &  $\mathbb R^{2,1} \times c(X^7)$ \\
 D3-brane & IIB & AdS$_5\times X^5$ &  $\mathbb R^{3,1} \times c(X^5)$ \\ 
 D1+D5-brane & IIB & AdS$_3\times X^3\times \mathbb T^4$ & $\mathbb R^{1,1} \times c(X^3)\times \mathbb T^4$ \\
 NS1+NS5-brane & IIA, B \& heterotic & AdS$_3\times X^3\times \mathbb T^4$ & $\mathbb R^{1,1} \times c(X^3)\times \mathbb
T^4$ \\
 \bottomrule
\end{tabular}
 \caption{Some brane solutions of 10- and 11-dimensional supergravities \cite{Acharya98}. Here $X^k$ is
a compact Einstein space of dimension $k$ and $c(X^k)$ denotes its metric cone, which is
Ricci-flat. The NS1-brane is also called a `fundamental string' (or `F1-string') \cite{DGHR90} and the NS1+NS5-brane system a `dyonic string' \cite{Duffdyonicstring}.}\label{tab:oldbranes}
\end{table}
 The solutions interpolate between a near horizon anti-de Sitter geometry AdS$_p\times X^k$ and $\mathbb
R^{p-2,1}\times c(X^k)$, where $X^k$ is an Einstein manifold and $c(X^k)$ its metric cone. They preserve some
supersymmetry if and only if $X^k$ carries a so-called geometric (real) Killing spinor, which is equivalent to the
existence of a parallel spinor on the cone. If $X^k$ equals a round sphere $S^k$ then the near horizon solutions
preserve the maximum possible amount of supersymmetry. Manifolds with geometric Killing spinors have been classified by
B\"ar \cite{Baer93}, besides the spheres there are only four types, listed in Table \ref{tab:Killingspinormfs}.
 \renewcommand{\arraystretch}{1.2}
  \begin{table}[H]\centering
 \begin{tabular}{ccc}  \toprule
   $X$ & $\dim X$ &  Killing spinors\\ \hline  
   nearly K\"ahler & 6 & (1,1) \\
   nearly parallel $G_2$ & 7 &  (1,0) \\
  Sasaki-Einstein & $4n-1$ &  (2,0) \\
  Sasaki-Einstein &  $4n+1$ & (1,1) \\
  3-Sasakian & $4n+3$ &  ($n+2,0$) \\
   $S^n$   & $n$ &  ($2^{[n/2]},2^{[n/2]})$\\
 \bottomrule
\end{tabular}
 \caption{The classification of manifolds with geometric real Killing spinors. The two numbers of Killing spinors are for opposite signs of the Killing constant.
}\label{tab:Killingspinormfs}
\end{table}
 There have been some indications that heterotic supergravity admits similar solutions with near horizon geometry
AdS$_3\times X^k\times \mathbb T^{7-k}$, for $2\leq k\leq 7$, obtainable from the fundamental strings of
\cite{DGHR90} by including $\alpha'$ corrections \cite{DabhMurthy07,LSS07,KLS07,Duff08}.
Such backgrounds have been constructed for $X = S^2$, based on a 5-dimensional black hole solution \cite{CDKL07}, and also for $X =
S^3$ \cite{Duff:gaugedyonicstring,LLOP99}, given by the NS1+NS5-brane system on Minkowski space $\mathbb R^{9,1}$, i.e.
a fundamental string inside an NS5-brane.

Fundamental strings can be constructed on much more general geometries than just flat space; one only needs a
non-compact Ricci-flat Riemannian manifold of dimension at most eight, equipped with a non-trivial harmonic function.
The fundamental string world-volume can then be identified with an orthogonal 2-dimensional Minkowski space. It requires
more work to generalize NS5-branes to curved geometries, but this is possible as well in the heterotic setting. The basic observation is that
NS5-branes in heterotic supergravity are associated to `point-like instantons', i.e. singular Yang-Mills fields whose
singular support is the brane world-volume \cite{Witten95,Kanno99}. By a theorem of Tian, this singular subspace is calibrated and of
codimension four (at least), as required for a 5-brane \cite{Tian00,DS09}. One can smear the brane by deforming the instanton
slightly, and hence obtain a so-called gauge 5-brane \cite{Strom90}, which is smooth. Based on higher-dimensional
instantons \cite{FN84,FN85,CGK85,IP92,ILPR09,HILP10,BILL10,HILP11}, several generalizations of the gauge 5-brane have
been constructed, both on Minkowski space \cite{HS90,Khuri93,GN95,Loginov08} and on Ricci-flat cones \cite{HN11}. All of these
gauge branes possess an NS5-brane limit as well, where the instanton acquires a singularity. 

Using these results, we construct supersymmetric NS1+NS5-brane systems in heterotic supergravity that interpolate
between a near horizon AdS$_3 \times X^k\times \mathbb T^{7-k} $-limit, and the vacuum solution $\mathbb R^{1,1}\times
c(X^k)\times \mathbb T^{7-k}$, for $k =3,5,6,7$. As above, $X^k$ is an arbitrary geometric Killing spinor manifold of
appropriate dimension. Our construction yields an arbitrary number of fundamental strings, but only a single NS5-brane,
unlike the old solution on $X=S^3$ \cite{LLOP99} which allows also for multiple 5-branes. Heterotic supergravity involves two gauge fields, one of them
is responsible for the NS5-brane, the other one can be used to form also a gauge anti-5-brane, without spoiling the asymptotic behaviour.

Our supergravity solutions for a system of fundamental strings in an NS5-brane naturally resolve an
interpretational difficulty of higher-dimensional instantons in string theory. The problem is that the singular support
of point-like instantons on Euclidean space or a cone in explicit examples is often not of codimension four, as
would be appropriate for an NS5-brane, so one might even think that they describe branes of lower dimension. This
would lead to divergent ADM masses, however, since the fall-off of the relevant functions in the solutions is
of order $1/r^2$, which gives finite ADM masses only for 5-branes \cite{Strom90,HS90,Khuri93,GN95}. We argue in favour of a
5-brane interpretation here; the dimension of the singular support depends
on a choice of partial compactification of an open cylinder $\mathbb R_{>0}\times X$, with the branes localized at the
boundary $\{r=0\}$. When we add fundamental strings the compactification comes out right automatically, with a boundary
component $\{0\}\times X$, and if $\dim X>3$ then the world-volume of the 5-brane intersects the boundary non-trivially.
 Another possibility in the case of NS5-branes only is a one-point compactification, which leads to a manifold
diffeomorphic to the cone $c(X)$. This is the conventional choice but gives rise to a brane world-volume of the wrong
dimension, since the intersection of the brane with the boundary has been shrunk to a point.

The amount of
supersymmetry preserved by our backgrounds depends only on $k$ and the four types of admissible geometries for $X^k$.
Contrary to the expectation expressed in \cite{DabhMurthy07,LSS07,KLS07,Duff08}, where (largely hypothetical)
backgrounds asymptotic to AdS$_3\times S^k\times \mathbb T^{7-k}$ are studied, we do not find any maximally
supersymmetric solutions. 
For instance, the AdS$_3\times S^3\times \mathbb T^4$ near horizon limit of the ordinary NS1+NS5-brane \cite{LLOP99}
preserves eight supersymmetries out of sixteen, and this is the maximum amount possible for our construction.
However, this result should not be too surprising, given that the fundamental strings and NS5-branes themselves do not
preserve maximal supersymmetry.
The solutions show the expected supersymmetry enhancement; the near horizon limits have constant dilaton and preserve
twice as much supersymmetry as the full solutions do. The results are summarized
in Table \ref{tab:SUSYpreserved}.
 \renewcommand{\arraystretch}{1.2}
  \begin{table}\centering
 \begin{tabular}{ccccc}  \toprule
  $\dim X$ & $X$ &  SUSYs & near horizon SUSYs & $\mathfrak{isom}$\\ \hline  
   3 & $S^3$  & 4 &8  & $\mathfrak{psu}(1,1|2)$\\
   5 & Sasaki-Einstein  & 2 & 4& $\mathfrak{osp}(2|2)$  \\
  6&  nearly K\"ahler   & 1 & 2 & $\mathfrak{osp}(1|2)$ \\
  7 &nearly parallel $G_2$ & 1& 2& $\mathfrak{osp}(1|2)$ \\
  7 &Sasaki-Einstein  & 2 &4& $\mathfrak{osp}(2|2)$ \\
  7 & 3-Sasakian & 3 & 6&  $\mathfrak{osp}(3|2)$ \\
 \bottomrule
\end{tabular}
 \caption{Amount of supersymmetry preserved by the full heterotic supergravity solutions and their near horizon geometries
AdS$_3\times X^k\times \mathbb T^{7-k}$. Only simply connected manifolds $X$ are considered. The last column gives
the super isometry algebra of the near horizon geometry, modulo purely bosonic algebras. The number of global supersymmetries coincides with the number of supersymmetries of the fundamental string on the cone over $X$, unless $X$ is a round sphere (cf. Table \ref{tab:fdtstringSUSY} below).}\label{tab:SUSYpreserved}
\end{table}

 Our heterotic supergravity solutions deviate from the other supergravity branes in another way. The metric on a Sasaki-Einstein or
a 3-Sasakian manifold admits a canonical one-parameter family of deformations away from the Einstein metric, and it
turns out that in the near horizon limit AdS$_3\times X^k\times \mathbb T^{7-k}$ the metric on $X^k$ is not Einstein, but a particular deformed metric. For nearly K\"ahler and nearly parallel $G_2$ manifolds the
near horizon geometry requires the Einstein metric, however. Let us illustrate this for the seven-sphere. The round
metric on $S^7$ is 3-Sasakian, and hence also Sasaki-Einstein and nearly parallel $G_2$. We can represent $S^7$ as the
total space of a U(1)-fibration over $\mathbb CP^3$
 \begin{equation}\label{S7:CP3-fibration}
    S^1\hookrightarrow S^7 \rightarrow \mathbb CP^3,
 \end{equation} 
 or as the total space of an $S^3 =\mbox{SU}(2)$-fibration over $S^4$ 
 \begin{equation}\label{S7:S4-fibration}
   S^3 \hookrightarrow S^7 \rightarrow S^4.
 \end{equation} 
 Viewed as a Sasaki-Einstein manifold, $S^7$ gives rise to a near horizon solution AdS$_3\times S^7$ preserving four
supersymmetries, where the metric on $S^7$ is obtained by a deformation of the round metric along the Hopf fibration
\eqref{S7:CP3-fibration}. Viewing $S^7$ as a 3-Sasakian manifold we obtain a solution preserving six
 supersymmetries, and the metric is obtained by deforming the round metric along the fibration \eqref{S7:S4-fibration}.
We can also
equip the round $S^7$ with a nearly parallel $G_2$-structure, and thus obtain a near horizon solution preserving only
two supersymmetries. Furthermore, every 7-dimensional 3-Sasakian manifold admits a second nearly parallel $G_2$-metric
among its family of deformations \cite{Friedr97}, giving rise to the squashed seven-sphere in our
case and leading to another supergravity background with two supersymmetries. Hence, we obtain four supergravity
solutions with asymptotic AdS$_3\times S^7$ geometries, where $S^7$ comes equipped with four different metrics, two of
them Einstein, two of them not. The other limit is flat $\mathbb R^{1,1}\times \mathbb R^8$ for all but the squashed
seven-sphere cases. The same reasoning applies to any other 7-dimensional 3-Sasakian manifold instead of $S^7$.

The asymptotic AdS$_3$ region of our supergravity solutions can be taken as an indication that they are holographic,
with a dual 2-dimensional conformal field theory (CFT). Although we do not perform a detailed study of holography in
this work, we present an obvious candidate for the CFT which has the right symmetries. It is simply the world-sheet sigma
model with target space the supergravity near horizon geometry. In particular, the near horizon super isometry algebras
are `heterotic' in the sense that they consist of a left-moving supersymmetric algebra and a right-moving bosonic
algebra. A holographic duality between the world-sheet CFT and the supergravity backgrounds would confirm the
interpretation of the geometries as `fundamental strings', but it is not clear how the 5-branes enter in this story.

The paper is organized as follows. We briefly review heterotic supergravity in Section~\ref{sec:HetSugra}, before we
discuss stabilizer groups of spinors in ten dimensions in Section~\ref{sec:stab}, which will be needed for the solutions
of the gravitino equation. In Section~\ref{sec:oldsolns} we review three heterotic BPS solutions which will be used in
Section~\ref{sec:oursol}, namely the gauge 5-branes, NS5-branes and fundamental strings.
Section~\ref{sec:oursol} contains the main result of the paper, i.e. the construction of new heterotic BPS backgrounds
which are shown to interpolate between an AdS$_3$ region and a Ricci-flat cone. We consider first the most general
setting with a gauge anti-5-brane present in Subsections \ref{sec:G2}--\ref{sec:3S}, and the simpler case of an
NS1+NS5-brane system only in \ref{ssec:F1+NS5}. Global properties and the relation to calibrated geometry are discussed
in \ref{ssec:Top+cycles}.
 For completeness' sake we also present the NS1 plus gauge 5-brane system,
which has a different asymptotic behaviour, in Subsection \ref{sec:gauge5brane}. The prototypical NS1+NS5-brane
asymptotic to
AdS$_3\times S^3\times \mathbb T^4$ is reviewed in Paragraph~\ref{sec:se_sol}; we consider $S^3$ as a 3-dimensional
Sasaki-Einstein manifold, and discuss a family of solutions for arbitrary 3-, 5- or 7-dimensional Sasaki-Einstein
manifolds.
The final Section~\ref{sec:Isoms} deals with isometries and holography.

\section{Heterotic supergravity}\label{sec:HetSugra}
 Heterotic supergravity consists of 10D $\mathcal N=1$ supergravity coupled to super-Yang-Mills. 
The ingredients are a 10-dimensional manifold $M$, equipped with a Lorentzian metric $g$, a 3-form $H$, scalar field $\phi$
and gauge connection $\nabla^A$, with gauge group SO(32) or $E_8\times E_8$. Denote by $F$ the curvature 2-form of
$\nabla^A$, and by
$\nabla^\pm$ the metric compatible connections on the tangent bundle of $M$ with torsion $\pm  H$, i.e. in terms of
connection coefficients
 \begin{equation}
   {\Gamma^\pm}^\mu_{\nu\lambda} = \Gamma^\mu_{\nu\lambda } \mp \frac 12 {H^\mu}_{\nu\lambda},
 \end{equation} 
 where $\Gamma^\mu_{\nu\lambda}$ are the coefficients of the Levi-Civita connection.
The BPS equations up to order $\alpha'$ are 
 \begin{equation}\label{BPSeqtns}
  \begin{aligned}{}
    \nabla^- \epsilon &=0, \\
    (\dd\phi-\sfrac 12 H)\cdot \epsilon & =0,\\
    F\cdot \epsilon &=0,
  \end{aligned}
 \end{equation} 
 for a Majorana-Weyl spinor $\epsilon$. The Clifford action of a $p$-form $\omega$ on a spinor $\epsilon$ is given by 
 \begin{equation}
   \omega \cdot \epsilon = \sfrac 1{p!} \omega_{\mu_1 \dots \mu_p} \gamma^{\mu_1 \dots \mu_p} \epsilon,
 \end{equation} 
 and we use the convention $\{\gamma^\mu,\gamma^\nu\} = 2g^{\mu\nu}$.
The equations of motion are
 \begin{equation} \label{eom}
\begin{aligned}{}
\text{Ric}_{\mu\nu} +2 (\nabla \dd\phi)_{\mu\nu} 
-\sfrac 14 H_{\kappa\lambda\mu}{H_{\nu}}^{\kappa\lambda}
+\frac {\alpha'}4 \Big[ \tilde R_{\mu \kappa\lambda\sigma} \tilde R_\nu^{\ \kappa\lambda\sigma} 
- \text{tr}\big(F_{\mu\kappa} {F_\nu}^\kappa\big)\Big]&=0, \\
\text{Scal} + 4 \Delta \phi  -4|\dd\phi|^2 -\sfrac 12| H|^2
+\frac {\alpha'} 4\text{tr}\Big[ |\tilde R|^2 - |F|^2 \Big]&=0 ,\\[2pt]
e^{2\phi }\dd\ast(e^{-2\phi}F) + A\wedge\ast F - \ast F\wedge A+\ast H\wedge F&=0, \\[4pt]
\dd\ast e^{-2\phi}H &=0. 
\end{aligned}
\end{equation}
 Here $|\omega|^2 = \frac 1{p!} \omega_{\mu_1 \dots \mu_p} \omega^{\mu_1\dots \mu_p}$ for a $p$-form $\omega$. The
dilaton equation has been used to bring the Einstein equation into a simpler form. Additionally, one has to impose the
Bianchi identity
 \begin{equation}\label{Bianchi}
   \dd H = \frac {\alpha'}4 \text{tr} \Big(\tilde R\wedge\tilde  R - F\wedge F\Big),
 \end{equation} 
 where `tr' is a positive-definite inner product on the gauge algebra, actually minus the ordinary trace over the tangent
space in our case. Here $\tilde R$ is the curvature form of a connection $\tilde \nabla$ on the tangent bundle, and there has
been some debate on the correct choice of $\tilde \nabla$. String theory appears to prefer the choice $\tilde
\nabla=\nabla^+$ \cite{BergdeRoo89,BeckerSethi09}, whereas a purely supergravity point of view seems to indicate that
$\tilde R$ must satisfy the instanton equation $\tilde R\cdot \epsilon=0$ \cite{Iv09}. Usually, both conditions cannot
be satisfied at the same time, a notable exception being the NS5-brane in flat space-time \cite{CHS91a}.

We will adopt the supergravity point of view, and impose the instanton condition on $\tilde R$. Then the BPS
equations together with the Bianchi identity and the time-like components of the field equations imply the remaining
components of the field equations \cite{GMPW02,Iv09}, which simplifies the calculations considerably and guarantees
that we get a consistent supergravity theory, independently of any string theory embedding. If one insists instead on $\tilde
R=R^+$, then the BPS equations and Bianchi identity only imply the field equations up to higher order corrections in
$\alpha'$, and one needs the full tower of stringy $\alpha'$-corrections to obtain a consistent supergravity
theory. It has also been argued, however, that the two approaches are equivalent via field redefinitions \cite{Hull86}, and indeed the near horizon limit of NS5-branes on Ricci-flat cones can be obtained in both settings, $\tilde R=R^+$ \cite{Noe11} and $\tilde R \cdot \epsilon=0$ \cite{HN11}.

Note that it is very natural in heterotic supergravity to include the first order $\alpha'$-corrections, since at
zeroth order the gauge field decouples, and one loses some of the massless modes of the corresponding string theory. On
the other hand, it is not entirely clear that a supergravity solution can be lifted to a full string background, since
solutions to the first order equations \eqref{eom} often depend explicitly on $\alpha'$, and higher order corrections 
potentially become large.

With our convention for $\tilde \nabla$ we can treat the two connections $\nabla^A$ and $\tilde \nabla$ on equal
footing; for a supersymmetric solution they both have to satisfy the instanton equation $F\cdot \epsilon= \tilde R\cdot
\epsilon=0$. In \cite{HN11} a 1-parameter family of instantons on the tangent bundle of the cone over a geometric
Killing spinor manifold $X$ was constructed, which interpolates between the Levi-Civita connection on the cone and the
pull-back of a canonical instanton connection $\nabla^P$ on $X$. In previous work on gauge solitonic branes the
connection $\tilde \nabla$ has always been identified with the Levi-Civita connection of the cone $c(X)$
\cite{Strom90,HS90,GN95,Loginov08,HN11}. In order to obtain the desired asymptotic behaviour we will instead identify the gauge connection $\nabla^A$
with the connection $\nabla^P$ on $X$, and choose $\tilde \nabla$ to be an interpolating instanton. The two conventions lead to opposite magnetic charges, and should be understood as brane and anti-brane solutions.

\section{Spinor stabilizers}\label{sec:stab}
 In Section \ref{sec:oursol} we will use the holonomy principle to solve the gravitino equation $\nabla^-\epsilon=0$, which tells us that the equation has $m$ solutions $\epsilon$ if and only if the holonomy group of $\nabla^-$ is contained in the joint stabilizer subgroup of $m$ spinors. The relevant stabilizer subgroups of Spin(9,1) are given in Table \ref{tab:stab}.
 \renewcommand{\arraystretch}{1.2}
  \begin{table}[H]\centering
 \begin{tabular}{cc}  \toprule
   $G$ & invariant spinors \\ \hline
   Spin(7$)\ltimes \mathbb R^8$ & 1 \\
   SU(4$)\ltimes \mathbb R^8$ & 2 \\
   Sp(2$)\ltimes \mathbb R^8$ & 3 \\
   \big(SU(2$)\times \mbox{SU}(2)\big)\ltimes \mathbb R^8 $ & 4 \\
   $\mathbb R^8$ & 8 \\
   $G_2$ & 2\\
   SU(3) & 4\\ 
   SU(2) & 8 \\
   $\{1\}$ & 16\\
 \bottomrule
\end{tabular}
\caption{Stabilizer subgroups $G$ of Spin(9,1) for a given number of Majorana-Weyl spinors with fixed chirality \cite{GLP05}.}\label{tab:stab}
\end{table}
  
Note that the stabilizer groups come in two flavours, compact ones and non-compact ones. Furthermore, whenever $G$ is a compact stabilizer group, then the non-compact group $G\ltimes \mathbb R^8$ also leaves some spinors invariant, since it is contained in a larger non-compact stabilizer. For instance, $G_2\ltimes \mathbb R^8\subset \mbox{Spin}(7)\ltimes\mathbb R^8$. For our heterotic supergravity backgrounds, non-compact stabilizers will be relevant. 

The non-compact subalgebra $\mathbb R^8$ of $\mathfrak{so}(9,1)$ is obtained as follows. Consider $\mathbb R^{9,1}$ with coordinates $x^\mu$, where $\mu=0,\dots,9$. For a matrix $X\in \mathfrak{so}(9,1)$ define $X_{\mu\nu} = \eta_{\mu\lambda}{X^\lambda}_\nu$, which is antisymmetric in its lower indices. The subalgebra $\mathbb R^8$ is defined by the equations
 \begin{equation}
   X_{a9} = X_{a0},\qquad X_{ab}=0,
 \end{equation} 
 for all $a,b=1,\dots,8$. A set of generators $\{I_a\}$ can be defined by
  \begin{equation}\label{R8-generators}
  \begin{aligned}{}
    {(I_a)^b }_9& = -  {(I_a)^9}_b ={\delta^b}_a, \\ 
    {(I_a)^b}_0 &= {(I_a)^0}_b ={\delta^b}_a .
  \end{aligned}
 \end{equation} 
 The non-compact stabilizer subgroups of Table \ref{tab:stab} are of the form $G\ltimes \mathbb R^8$, where $G$ is a
subgroup of SO(8). Let us also introduce a generator $Z$ of the algebra $\mathfrak{so}(1,1)\subset \mathfrak{so}(9,1)$
orthogonal to $\mathfrak{so}(8)$, as
   \begin{equation}\label{Zgenerator}
    {Z^0}_9 = {Z^9}_0 =2.
  \end{equation} 
 The generator $Z$ commutes with the subalgebra $\mathfrak{so}(8)$ of $\mathfrak{so}(9,1)$, and leaves $\mathbb R^8$
invariant 
\begin{equation}\label{commut_ZIa}
  [Z,I_a] =-2 I_a.
\end{equation} 
 An $\mathbb R^8$-invariant spinor $\epsilon$ is characterized by the projection property
\begin{equation}\label{spinorprojectionprop}
 \gamma^0 \epsilon = - \gamma^9 \epsilon \; .
\end{equation}
As an element of $\mathfrak{spin}(9,1)$ we have $Z = - \gamma^0\gamma^9$, and \eqref{spinorprojectionprop} shows that
\begin{equation}\label{Zonspinor}
 Z\epsilon = -\epsilon \; .
\end{equation}

\section{Old solutions}\label{sec:oldsolns}
   We briefly review the gauge solitonic brane solutions of \cite{Strom90,HS90,GN95,Loginov08,HN11} together with their NS5-brane limit \cite{LLOP99}, and the fundamental string of \cite{DGHR90}, which will be ingredients of our heterotic supergravity solutions to be developed in the following section. 
\subsection{Gauge solitonic branes and NS5-branes}\label{sec:gaugesolbrane}
 The gauge solitonic 5-branes can be defined on a manifold of the form
 \begin{equation}
  M= \mathbb R^{1 ,1}\times \mathbb T^{7-k}\times \mathbb R\times X^k, 
 \end{equation} 
 where the fields depend trivially on the $\mathbb R^{1,1} \times \mathbb T^{7-k}$ factor. The
manifold $X^k$ carries a so-called geometric real Killing spinor, i.e. a spinor $\epsilon$ which satisfies
 \begin{equation}
   \Big(\nabla_\mu - \frac i2 \gamma_\mu \Big) \epsilon =0.
 \end{equation} 
  Here $\nabla$ denotes the Levi-Civita (or spin) connection. The geometric Killing spinor equation implies that $X^k$ is Einstein, with Einstein constant $k-1$. The metric on space-time is chosen in the form
 \begin{equation}\label{gaugebranemetric}
   g = -\dd t^2+\dd x^2+ g_{\mathbb T^{7-k}}+ e^{2f(\tau)} \big(\dd\tau^2 + g^k),
 \end{equation} 
 where $\tau$ is the linear coordinate on $\mathbb R$ and $g^k$ is a possibly $\tau$-dependent metric on $X^k$. Every
geometric Killing spinor manifold, except possibly the even-dimensional spheres in dimension not equal to six, comes
equipped with a reduction of the structure group SO($k$) to some subgroup $K$, a $K$-invariant 3-form $P$, as well as a
connection $\nabla^P$ with torsion proportional to $P$ and holonomy group contained in $K$. Furthermore, the cone
$c(X^k)$ with metric $\dd r^2 + r^2 g^k$ is Ricci-flat and carries an integrable reduction of the structure group SO($k+1)$
to a subgroup $G$. See Table \ref{tab:Killingmfs} for the groups $K$ and $G$ that occur, and
\cite{BFGK91,Baer93,Boyer07,HN11} for more details on the geometry of manifolds with geometric Killing spinors.
 
 \renewcommand{\arraystretch}{1.2}
  \begin{table}[H]\centering
 \begin{tabular}{cccc}  \toprule
  $\dim X$ & $X$ &  $K$ & $G$ \\ \hline  
  6&  nearly K\"ahler   & SU(3) & $G_2$\\
  7 &nearly parallel $G_2$ & $G_2$ & Spin(7)  \\
  $2n+1$ &Sasaki-Einstein  & SU($n$) & SU($n+1$)  \\
  $4n+3$ & 3-Sasakian & Sp$(n)$ & Sp$(n+1)$ \\
 $ n$ & $S^n$ & SO($n$) & $\{ 1\}$ \\
 \bottomrule
\end{tabular}
 \caption{B\"ar's classification of manifolds with geometric real Killing spinors \cite{Baer93}. $K$ is the structure group of $X$ but does not coincide with the holonomy group of its Levi-Civita connection, and $G$ is the holonomy group of $\nabla^c$, the Levi-Civita connection on the cone $c(X)$.}\label{tab:Killingmfs} 
\end{table}

 The solution of the gravitino equation $\nabla^-\epsilon=0$ is particularly important. A simple choice for the connection $\nabla^-$ would be the canonical connection $\nabla^P$ on $X^k$, since it is known to have reduced holonomy. There is some more freedom however. In \cite{HN11} a bundle map 
 \begin{equation}
   \rho: TX \rightarrow \mbox{End}\big(T(\mathbb R \times X)\big)
 \end{equation} 
 was constructed, whose image was shown to lie in the orthogonal complement of the subalgebra $\mathfrak k \subset
\mathfrak g$. Denote by $\{I_a\}$ a local basis of vector fields on $X$, and by $e^a$ the dual 1-forms. Then $e^a\otimes
\rho(I_a)$ is a globally defined section of the bundle $T^*(\mathbb R\times X) \otimes \mbox{End}(T(\mathbb R
\times X))$, which we denote simply by $e^a I_a$, and the connection $\nabla^-$ is constructed via the ansatz
 \begin{equation}\label{nabla-ansatz_old}
   \nabla^- = \nabla^ P + s(\tau) e^a I_a \; ,
 \end{equation} 
for some function $s(\tau)$ constrained by the requirement that the torsion of $\nabla^-$ be totally antisymmetric. 
 By construction, its holonomy group is contained in $G$, hence it has a parallel spinor. In Section~\ref{sec:oursol}, we will use a similar ansatz
for the connection $\nabla^-$, but allow for additional terms compatible with the larger holonomy group $G\ltimes
\mathbb R^8$. 
 The same ansatz 
 \begin{equation}\label{instansatzgeneric}
   \nabla(\psi)= \nabla^ P + \psi(\tau) e^a I_a
 \end{equation} 
 was chosen for the gauge connection $\nabla^A$ in \cite{HN11}, and the instanton (or
gaugino) equation reduces to a first order differential equation for $\psi$:
 \begin{equation}\label{insteqgeneric}
    \dot \psi = 2\psi(\psi-1)
 \end{equation} 
 for nearly K\"ahler and nearly parallel $G_2$ manifolds or $S^3$, where $\dot \psi =\partial_\tau\psi$.
In the case of a Sasakian manifold there are actually two independent sections that can be added to $\nabla^P$, and this
leads to slightly more complicated instanton equations. For a 3-Sasakian manifold they were solved analytically in
\cite{HN11}, but only numerically for a Sasaki-Einstein manifold. 

 The instanton equation \eqref{insteqgeneric} has two fixed points $\psi=0$ and $\psi=1$, corresponding to $\nabla^P$ and $\nabla^c$, the Levi-Civita connection on the cone. For $X^k=S^k$ the cone is flat Euclidean space, so the connection $\nabla^c$ has vanishing instanton charge. The other limit is more subtle; let us concentrate on the case $X=S^3$. Then the instanton number is proportional to tr$\int_{\mathbb R^4} F\wedge F$. Formally one finds that
 \begin{equation}
  \begin{aligned}{}
     \int\mbox{tr}( F\wedge F) & = -12\mbox{Vol}(S^3) \int_{-\infty}^\infty \dot \psi \psi(\psi-1) \dd\tau \\
       &= 2\mbox{Vol}(S^3) \big[3\psi^2 - 2\psi^3 \big] \Big|^{\tau = \infty}_{\tau=-\infty},
  \end{aligned}
 \end{equation} 
 and plugging in $\psi=0$ or $\psi=1$ gives rise to vanishing instanton number. A more careful analysis involves the general solution of \eqref{insteqgeneric}, which is given in terms of a radial variable $r=e^\tau$ by
 \begin{equation}\label{instgensol}
   \psi(r) = \frac {\rho^2}{\rho^2+ r^2},
 \end{equation} 
 where the parameter $\rho \in [0,\infty]$. For $\rho \neq 0,\infty$ the solution interpolates between zero and one, and the instanton number is proportional to
 \begin{equation}
   \int\mbox{tr}( F\wedge F)  =  2\mbox{Vol}(S^3) ,
 \end{equation} 
 independently of $\rho$. Now it turns out that the integrand $\mbox{tr}( F\wedge F)$ divided by the Euclidean volume
form for $\mathbb R^4$ becomes more and more concentrated around $r=0$ as $\rho \rightarrow 0$. Hence, we should
interpret the limiting case $\psi=0$ in a distributional way as a point-like instanton \cite{Witten95,NekrSchw98}, and
assign to it charge 1, like for the generic solution \eqref{instgensol}. The other limiting case $\psi=1$ is perfectly
regular on the other hand, and is rightly assigned instanton charge zero. What is the supergravity interpretation of the
different instantons? First of all, there are two gauge fields, $\nabla^A$ and $\tilde \nabla$, which we choose both to
be of the form \eqref{instansatzgeneric}; $\tilde \nabla = \nabla(\psi_1)$ and $\nabla^A= \nabla(\psi_2)$. For
$\psi_1=1$ and $\psi_2$ generic we obtain Strominger's gauge solitonic 5-brane \cite{Strom90}, which is a regular
supergravity solution. In the limit $\psi_2\rightarrow 0$ the solution develops a singularity at $r=0$ and it becomes an
NS5-brane \cite{CHS91a,LLOP99}. Since $\tilde \nabla$ leads to opposite charge than $\nabla^A$, we will interpret the
case $\psi_1\neq 1$ as an anti-brane. This is summarized in Table \ref{tab:braneinterpretation}.
  \renewcommand{\arraystretch}{1.2}
  \begin{table}[H]\centering
 \begin{tabular}{ccc}  \toprule
   $(\rho_1,\rho_2)$ & brane system & total brane charge \\ \hline  
    ($\infty$,0) & NS5-brane  & 1 \\
    ($\infty$,$\rho_2$) & gauge 5-brane & 1\\
    ($\rho_1$,$\rho_2$) & gauge anti-5-brane + gauge 5-brane & 0\\
    ($\rho_1,0$) & gauge anti-5-brane + NS5-brane &0 \\
 \bottomrule
\end{tabular}
 \caption{Brane configurations for different choices of the gauge fields $\tilde \nabla$ and $\nabla^A$. We always assume that $\rho_1>\rho_2$, which is required for a non-singular metric in the region $0<r<\infty$. In the limit $\rho_1=\rho_2$ the $\alpha'$-corrections vanish and we are left with the Ricci-flat cone solution.
Below the headline $\rho_1,\rho_2$ denote generic values, i.e. $\rho_1,\rho_2\neq 0,\infty$. The special values $\rho=0,\infty$ correspond to $\psi=0,1$, respectively. The magnetic or brane charge contribution of $\rho_2<\infty$ is 1, whereas $\rho_1<\infty$ contributes $-1$. There is no charge contribution in the limiting case $\rho=\infty$.}\label{tab:braneinterpretation} 
\end{table}
The discussion for the case $X=S^3$ applies to higher dimensions as well; the limiting connection $\nabla(\psi=0)$ is a singular charge one instanton, like the smooth interpolating solutions for generic $\psi$, whereas $\nabla^c$ has vanishing instanton charge. The supergravity solution corresponding to $\psi=0$ is an NS5-brane, whereas the interpolating instantons
give rise to smooth gauge 5-branes, which can be viewed as smeared NS5-branes \cite{Witten95}.

The ansatz for the gauge field we have presented here leads to instanton number plus/minus one, and hence to a single
brane, or a brane-anti-brane system. Explicit multi-instanton and multi-brane solutions are known only for
$X=S^3,S^7,S^8$, or $c(X)=\mathbb R^4,\mathbb R^7,\mathbb R^8$ \cite{Strom90,Loginov05}. The amount of supersymmetry
preserved by a gauge 5-brane or an NS5-brane coincides with the supersymmetries of the Ricci-flat cone, except when the
cone is flat. In the latter case we need to fix a Spin(7), SU(4), Sp(2),
$G_2$, Sp(2), SU(3) or SU(2)-structure on $\mathbb R^{9,1}$ to define the brane, and the amount of supersymmetry is
given by the number of invariant spinors, according to Table \ref{tab:stab}. In the near horizon limit of the NS5-brane
supersymmetry enhancement takes place; we have $\nabla^A=\nabla^-= \nabla^P$, so the relevant holonomy group reduces to
$K$. On the other hand, $\tilde \nabla = \nabla^c$ has holonomy group $G$, so unless $X$ is a round sphere the amount
of supersymmetry preserved depends on whether we require all supersymmetry generators to be annihilated by $\tilde R$
as well, or not.

 The $\alpha'$ corrections in the heterotic supergravity equations are essential for the gauge solitonic branes, in particular the 3-form $H$ is not closed in general and the modified Bianchi identity
 \begin{equation}
   \dd H = \frac {\alpha'}4 \mbox{tr}\Big(\tilde R\wedge \tilde R - F\wedge F\Big)
 \end{equation} 
 plays an important role. Suppose then that a solution with maximal supersymmetry exists, which implies that the spinor bundle is trivialized by a set of globally defined $\nabla^-$-parallel spinors $\{\epsilon_i\}$. The common stabilizer subgroup of the spinors is the trivial group, and the gaugino equation $F\cdot \epsilon_i = 0$ and the requirement $\tilde R \cdot \epsilon_i=0$ imply that $F=\tilde R=0$. But then the (first) $\alpha'$-corrections to the equations vanish, and we end up with a solution to the zeroth order equations. Hence, a maximally supersymmetric heterotic string background cannot receive $\alpha'$-corrections.

\subsection{The fundamental string}\label{sec:fundamentalstring}
 Here space-time is of the form $\mathbb R^{1,1} \times M^{k+1} \times \mathbb T^{7-k}$, with $M^{k+1}$ a non-compact
Ricci-flat manifold. The fields are\footnote{We give the fields in the string
frame instead of the Einstein frame and
remark that both conventions are used in the literature on supergravity
solutions. The main difference is the way in which the harmonic function
appears in the metric, which is important to keep in mind when comparing
formul\ae\ in different papers.}
 \begin{equation}\label{F1}
   \begin{aligned}{}
      g&= h^{-1}(-\dd t^2 +\dd x^2) + g^{k+1} + g_{\mathbb{T}^{7-k}}, \\
      H&= \dd h^{-1} \wedge \dd t\wedge \dd x, \\ 
     e^{2(\phi-\phi_0)}&= h^{-1},
   \end{aligned}  
 \end{equation} 
 with $h$ a harmonic function on $M^{k+1}$. In addition, $h$ satisfies a quantization condition, which is essential
for the interpretation of the solution as a superposition of classical strings \cite{DGHR90}. The gauge fields $\tilde \nabla$ and $\nabla^A$ are both given by the Levi-Civita connection on $M$, so that all first order corrections in $\alpha'$ vanish.
 The amount of
supersymmetry preserved by the fundamental string solution depends on the amount of supersymmetry of the Ricci-flat
solution $\mathbb R^{1,1} \times M^{k+1} \times \mathbb T^{7-k}$ with vanishing fluxes. Suppose that the spinor
$\epsilon$ is parallel on the Ricci-flat geometry. Then it gives rise to a solution of the BPS equations for the
fundamental string if and only if it has the projection property
 \begin{equation}
   (\dd t\wedge \dd x)\cdot\epsilon= \epsilon,
 \end{equation} 
 which is equivalent to \eqref{spinorprojectionprop} and hence to $\epsilon$ being $\mathbb R^8$-invariant.  
 This implies in particular that maximal supersymmetry does not occur for the fundamental string. If $M$ has
holonomy group $G$ then the amount of supersymmetry preserved by the fundamental string equals the number of spinors
invariant under $G\ltimes \mathbb R^8$:
 \renewcommand{\arraystretch}{1.2}
  \begin{table}[H]\centering
 \begin{tabular}{cccc}  \toprule
   $\dim M$ & Hol$(M)$ & SUSYs (Ricci-flat) & SUSYs (string) \\ \hline
     8   &  Spin(7) & 1  & 1 \\
     8  & SU(4) & 2 & 2 \\
     8 & Sp(2) & 3 & 3\\
     8 & SU(2$)\times \mbox{SU}(2)$ & 4 & 4 \\	
     7 & $G_2$ & 2& 1 \\
     6 & SU(3) & 4 & 2 \\
     4 & SU(2) & 8 & 4\\
     $k+1$ & $\{1\}$ &   16 & 8 \\
 \bottomrule
\end{tabular}
 \caption{Amount of supersymmetry preserved by the Ricci-flat solution $\mathbb R^{1,1} \times M^{k+1} \times \mathbb T^{7-k}$ without fluxes, and a fundamental string on the same geometry. Hol$(M)$ denotes the holonomy group of $M$.}\label{tab:fdtstringSUSY} 
\end{table}
As an example consider the case $M^{k+1}=c(X^k)$ with metric $\dd r^2 + r^2 g^k$, the cone over a
geometric Killing spinor manifold $X^k$. Then one can write down an explicit solution for $h$:
 \begin{equation}
   h(r) = a^{-2}+\frac {Q_e} {r^{k-1}} \qquad (k\geq 3),
 \end{equation} 
 where $a$ and $Q_e$ are constants, and the electric charge $Q_e$ assumes a discrete set of values. It is proportional to $(\alpha')^{\frac{k-1}2} N$, with $N$ integer, but since the $\alpha'$ dependence does not follow from the supergravity equations we will not write it explicitly. $N$ is the number of strings. 
 For later convenience we collect the asymptotic behaviour of the fields as $r\to 0$ and $r\to \infty$.
\begin{align}
 &\begin{aligned}\label{F1r=0}
  g&= h^{-1} (-\dd t^2+ \dd x^2) + \dd r^2 + r^2 g^k  + g_{\mathbb T^{7-k}} \; , \\
  H&= \dd h^{-1} \wedge \dd t \wedge \dd x \; ,\qquad e^{2(\phi-\phi_0)} = h^{-1}  \; , \qquad h^{-1} = \frac{r^{k-1}}{Q_e} 
 \end{aligned}
 &&\left.\vphantom{\begin{pmatrix}a\\b\\c\end{pmatrix}}\right\} \qquad\text{as $r\to 0$, and} \\
 &\begin{aligned}\label{F1r=infty}
  g&= a^2 (-\dd t^2+ \dd x^2) + \dd r^2 + r^2 g^k + g_{\mathbb T^{7-k}} \; , \\
  H&= 0 \; ,\qquad e^{2(\phi-\phi_0)} = a^2   \vphantom{h^{-1} = \frac{\lambda}{Q_e} r^5}
 \end{aligned}
 &&\left.\vphantom{\begin{pmatrix}a\\b\\c\end{pmatrix}}\right\} \qquad\text{as $r\to\infty$}.
\end{align} 
 The solution interpolates between a warped product $\mathbb R^{1,1} \rtimes_{r^{k-1}} c(X^k) \times \mathbb{T}^{7-k}$ for $r\to 0$ and the vacuum solution
$\mathbb R^{1,1} \times c(X^k) \times \mathbb{T}^{7-k}$ for $r\to\infty$. At $r = 0$ the metric is singular.

\section{Asymptotically AdS\texorpdfstring{$_3$}{3} solutions}\label{sec:oursol} 
 In this section we will superpose the fundamental string solution and an NS5-brane to obtain new solutions of the heterotic BPS equations
\eqref{BPSeqtns} and the Bianchi identity \eqref{Bianchi}, as well as the time-like components of the field equations
\eqref{eom}. The four types of geometric Killing spinor manifolds are treated separately, but in all cases we find a near horizon AdS$_3$-region and a Ricci-flat cone in another limit. To begin with we consider the more general setting of fundamental strings, an NS5-brane and a gauge anti-5-brane, which leads to the same asymptotics, and later also treat the case of fundamental strings with a gauge 5-brane. In the latter case the near horizon AdS$_3$ disappears; the asymptotic solutions coincide with those of the fundamental string.

\subsection{Nearly parallel \texorpdfstring{$G_2$}{G2}}\label{sec:G2}
 Let $X^7$ be a 7-dimensional nearly parallel $G_2$ manifold.
We make an ansatz for the space-time manifold in the form
 \begin{equation}
   M= \mathbb R^ {1,1} \times \mathbb R \times X^7.
 \end{equation} 
  The metric is chosen as
 \begin{equation}
  g  = h^{-1}(\tau) (-\dd t^2 +\dd x^2 ) + e^{2f(\tau)}(\dd \tau^2 + g^7),
 \end{equation} 
 where $t$ and $x$ are coordinates on $\mathbb R^{1,1}$, and $\tau$ parametrizes the remaining $\mathbb R$-factor. By
$e^a$, for $1\leq a\leq 7$, we denote a basis of 1-forms on $X^7$, and $e^8:=\dd\tau$. It is useful to introduce the shorthand notation $e^{a_1 \ldots a_n} := e^{a_1} \wedge \cdots \wedge e^{a_n}$. The $G_2$-invariant 3-form $P$ on $X^7$ is then normalized such that
\begin{equation}
 P = e^{123} + e^{145} - e^{167} + e^{246} + e^{257} + e^{347} - e^{356} \; .
\end{equation}
It satisfies $\dd P = 4 \ast P$ reflecting the fact that $X^7$ is a 7-dimensional nearly parallel $G_2$ manifold.
\subsubsection{Gravitino equation}
 We make an ansatz for the connection $\nabla^-$ in the form  
 \begin{equation}\label{G2nabla-ansatz}
   \nabla^- = \nabla^ P + s(\tau) e^aI_a + \zeta(\tau) \dd t\, I_8 + \xi(\tau) \dd x\, I_8 + \alpha(\tau) \dd \tau \, Z,
 \end{equation} 
 where $\nabla^P$ is the canonical $G_2$-connection on $X^7$, the $I_a$ are generators of the orthogonal complement of
$\mathfrak{g}_2$ in $\mathfrak {spin}(7)$ as explained in Section \ref{sec:gaugesolbrane}, $I_8$ is one of the generators
\eqref{R8-generators} corresponding to the $\tau$-direction, and $Z$ is the $\mathfrak{so}(1,1)$-generator
\eqref{Zgenerator}. Let us introduce the orthonormal basis
  \begin{equation}
   \begin{aligned}{}
     \sigma^0 &= h^{-1/2} \dd t ,\qquad \sigma^9 = h^{-1/2} \dd x ,\\
      \sigma^8 &= e^f\dd\tau ,\qquad \quad\sigma^a =e^f e^a.
   \end{aligned}
 \end{equation} 
 Using the Cartan structure equation $T^\mu = \dd\sigma^\mu + {}^-{\Gamma^\mu}_\nu \wedge \sigma^\nu$ we can calculate
the
torsion of $\nabla^-$; the $T^a$-components from \cite{HN11} are unchanged, whereas we find additionally
 \begin{equation}
 \label{npG2:Ads3-torsion}
  \begin{aligned}{}
    T^0 &= (\partial_\tau h^{-1/2}-\zeta e^f) \dd\tau \wedge \dd t + (2\alpha h^{-1/2} -\xi e^f) \dd\tau \wedge \dd x,
\\
    T^9 &= (\partial_\tau h^{-1/2} + \xi e^f) \dd\tau \wedge \dd x + (2\alpha  h^{-1/2} + \zeta e^f) \dd\tau \wedge \dd t,
\\
    T^8 &= (\zeta-\xi)  h^{-1/2} \dd t \wedge \dd x.
  \end{aligned}
 \end{equation} 
 Since the torsion of $\nabla^-$ has to be totally antisymmetric, i.e. $T^\mu = \sigma^\mu \lrcorner H$ for some 3-form
$H$, we have to impose the conditions
  \begin{equation}\label{G2-Hsol}
  \begin{aligned}{}
    \zeta = -\xi &= e^{-f}\partial_\tau h^{-1/2} ,\qquad    4 \alpha = -\partial_\tau \log(h),\qquad s=\dot f ,\\
    H&= \dd h^{-1} \wedge \dd t \wedge \dd x -\frac 23 (\dot f-1) e^{2f} P,
  \end{aligned}
 \end{equation} 
 where $\dot f =\partial_\tau f$. To show
that $\nabla^-$ has holonomy group Spin(7$)\ltimes \mathbb R^8$ and hence exactly one parallel spinor we perform a gauge
transformation to eliminate the $Z$-term, using \eqref{commut_ZIa}:
\begin{equation}
  e^{-\frac 14 \log(h) Z}( \nabla^-) e^{\frac 14 \log(h) Z} = \nabla^P + \dot f e^aI_a -\sfrac 12  e^{-f}
\partial_\tau \log(h)(\dd t -\dd x) I_8.
 \end{equation} 
 In this form the Spin(7)$\ltimes \mathbb R^8$-holonomy becomes manifest. 

\subsubsection{Dilatino equation}
 The action of the 3-form $H$, as determined above, on the Spin($7)\ltimes \mathbb R^8$-invariant spinor $\epsilon$ is
given by
 \begin{equation}
   H \cdot \epsilon =\Big(- \partial_\tau \log(h) + \frac {14}3(\dot f-1) \Big) \dd\tau \cdot \epsilon.
 \end{equation}
 Hence, the dilatino equation $(\dd\phi - \frac 12 H)\cdot \epsilon=0$ is solved by 
 \begin{equation}\label{G2-phisol}
   \phi (\tau ) =\phi_0  -\sfrac 12 \log(h)   + \sfrac 73 (f-\tau).
 \end{equation} 

\subsubsection{Gaugino equation}
 The gaugino equation requires the gauge field to be a Spin(7)$\ltimes \mathbb R^8$-instanton, and we also impose this
condition on the connection $\tilde \nabla$. A connection of the form
 \begin{equation}\label{G2-gaugefieldansatz}
  \nabla^P + \psi(\tau) e^a I_a,
 \end{equation} 
 solves the Spin(7)-instanton equation if and only if $\psi$ satisfies~\cite{HN11}
 \begin{equation}\label{G2-psiinsteq}
   \dot \psi = 2\psi(\psi-1).
 \end{equation} 
 Besides the two fixed points $\psi=0$ and $\psi=1$ which correspond to the canonical connection $\nabla^P$ and the
Levi-Civita connection of the cone $c(X^7)$ respectively, there is an interpolating solution
 \begin{equation}
  \psi(\tau ) = \Big(1+ e^{2(\tau-\tau_0)}\Big)^{-1}.
 \end{equation}
 Denote the curvature form of \eqref{G2-gaugefieldansatz} by $\mathcal{F}(\psi)$. We put $\tilde R = \mathcal{F}(\psi_1)$ and
$F=\mathcal{F}(\psi_2)$, and later we will make the choice $\psi_2=0$.

\subsubsection{Bianchi identity}
 Since the term $\dd h^{-1} \wedge \dd t\wedge \dd x$ in $H$ is closed, the Bianchi identity essentially reduces to the
same equation as in \cite{HN11}:
\begin{equation}\label{G2-Bianchi1storder}
  (1-\dot f) e^{2f}  = \frac{\alpha'}4 \big(\psi_1 ^2 - \psi_1 \dot \psi_1 - \psi_2^2 + \psi_2\dot \psi_2 \big).
\end{equation} 

\subsubsection{Field equations}
 Besides the BPS equations and the Bianchi identity we also have to solve the time-like components of the field
equations. The other components of the field equations are then satisfied as well.
 Due to our special ansatz the $t$-component of the Yang-Mills equation and the mixed $(t\mu)$-component for
$\mu\neq t$ of the Einstein equation are trivially satisfied. It remains to consider the $t$-component of the $B$-field
equation and the $(tt)$-component of the Einstein equation.  
 For the former we calculate
 \begin{equation}
   \dd *e^{-2(\phi-\phi_0)} H = \dd \left[ \dot h\, \exp\Big(\sfrac 43f + \sfrac{ 14}3\tau\Big)\mbox{Vol}^7 \right],
 \end{equation} 
 where Vol$^7$ denotes the volume form of the nearly parallel $G_2$-metric on $X^7$.
 The $B$-field equation becomes
 \begin{equation}\label{G2-Bfieldeq}
   \partial_\tau \Big[\dot h\, \exp\Big(\sfrac 43f + \sfrac{ 14}3\tau\Big)\Big]=0,
 \end{equation}
 and this turns out to coincide with the ($tt$)-component of the Einstein equation.

\subsubsection{Solution}\label{sec:g2_sol}
 We already solved the gravitino, dilatino and gaugino equations, and found the following result for the metric, 3-form
and dilaton:
 \begin{equation}\label{g2_10dsol}
  \begin{aligned}{}
      g  &= h^{-1}(\tau) (-\dd t^2 +\dd x^2 ) + e^{2f(\tau)}(\dd \tau^2 + g^7), \\
      H&= \dd h^{-1} \wedge \dd t \wedge \dd x -\frac 23 (\dot f-1) e^{2f} P ,\\
     \phi (\tau ) &=\phi_0  -\sfrac 12 \log(h)   + \sfrac 73 (f-\tau).
  \end{aligned}
 \end{equation} 
  It remains to solve the $B$-field equation \eqref{G2-Bfieldeq} and the Bianchi identity \eqref{G2-Bianchi1storder}.
The former can be integrated to 
 \begin{equation}
   \dot  h = - 6Q_e \exp\Big(-\sfrac 43 f - \sfrac {14}3 \tau \Big),
  \end{equation}  hence a solution is given by
 \begin{equation}\label{g2_h_int}
 h(\tau) = a^{-2} + 6Q_e \int _\tau^{\infty} \exp\Big(-\sfrac 43 f (\theta)- \sfrac {14}3 \theta \Big) \dd\theta  .
 \end{equation} 
  In the case of a cone metric, $f(\tau)=\tau$, this reduces to a harmonic function
 \begin{equation}
   h(r) = a^{-2} + \frac{Q_e}{r^6},
 \end{equation} 
 written in terms of a radial coordinate $r=e^\tau$, and we recover the fundamental string of Section~\ref{sec:fundamentalstring}.

The connections  $\tilde \nabla$ and $\nabla^A$ are constructed by the ansatz \eqref{G2-gaugefieldansatz}, and depend
on functions $\psi_1(\tau)$ and $\psi_2(\tau)$, respectively. In order to obtain an NS5-brane, we set $\psi_2=0$. In most of the literature only the case $\psi_1=1$ is considered \cite{Strom90,HS90,GN95,Loginov08,HN11}, but here we keep $\psi_1$ generic, thus allowing for a gauge anti-5-brane as well, and treat the limiting case $\psi=1$ separately in Subsection \ref{ssec:F1+NS5}.
 \begin{equation}
   \psi_1 = \frac {\rho^2}{\rho^2+ r^2}, \qquad \psi_2= 0,
 \end{equation} 
 where $\rho$ is a constant.
 Then $\nabla^A = \nabla^P$, with $\nabla^P$ being the canonical $G_2$-connection on $X^7$. The Bianchi identity has been solved in
\cite{HN11}
\begin{equation}
 \begin{aligned}{}
      e^{2f} &= \lambda^2 r^2 + \frac {\alpha'}4 (\psi_1^2-\psi^2_2) \\ 
 &=\lambda^2 r^2+\frac {\alpha'}4\, \frac{\rho^4}{(\rho^2+ r^2)^2}   ,
 \end{aligned}
\end{equation} 
 for some constant $\lambda$. We note in passing that in the special case $\psi_1 = \psi_2$, one recovers the fundamental string without $\alpha'$-corrections as presented in Section~\ref{sec:fundamentalstring}.

\paragraph{Limit $r\rightarrow \infty$.} 
 In this limit we obtain the Ricci-flat cone $\mathbb R^{1,1} \times c(X^7)$:
 \begin{equation}\label{g2_sol_r_infty}
  \begin{aligned}{}
       g&= a^2 (-\dd t^2+ \dd x^2) + \lambda^2 (\dd r^2 + r^2 g^7 ), \\
        H&= 0,\qquad e^{2(\phi-\phi_0)} = a^2 \lambda^{\frac{14}3} .
  \end{aligned}
 \end{equation}

\paragraph{Limit $r\rightarrow 0$.} It is convenient to substitute $s^2 := h^{-1}= \frac 7{9Q_e}(\frac{\alpha'}4r^7)^{2/3}$ in this limit. Then
the fields read
 \begin{equation}\label{g2_sol_r_0}
  \begin{aligned}{}
    g&=  s^2\big(-\dd t^2+\dd x^2 \big) + \alpha'\Big(\frac 3{14}\Big)^2\, \frac{\dd s^2}{s^2} + \frac{\alpha'}{4}  g^7,\\
    H&= {\dd (s^2)} \wedge \dd t\wedge \dd x + \frac {\alpha'}6 \, P, \\
   e^{2(\phi-\phi_0)} &= \frac 7{9Q_e} \left(\frac{\alpha'}4\right)^3 .
  \end{aligned}
 \end{equation} 
 In particular, the dilaton is constant and it becomes small for a large number of strings. Hence we can trust the
supergravity approximation for large $Q_e$, a situation familiar from the other brane solutions \cite{Mald07}. The
metric describes a direct product AdS$_3\times X^7$, where the length scales of both AdS$_3$ and $X^7$ are of order
$\sqrt{\alpha'}$. Heterotic string backgrounds of the form AdS$_3$ times a nearly parallel $G_2$-manifold have been
anticipated in \cite{KO09}, where it was shown that they solve the gravitino and dilatino equations. The term $\dot
fe^aI_a$ appearing in the connection $\nabla^-$ \eqref{G2nabla-ansatz} vanishes in the limit $r\rightarrow 0$, hence the
holonomy reduces to $G_2\ltimes \mathbb R^8$. However, a simple calculation shows that the $\mathbb R^8$-component of
the curvature vanishes, and the holonomy in fact reduces to $G_2$. According to Table \ref{tab:stab} this means that
another parallel spinor emerges. One can check that it also satisfies the dilatino equation, using an explicit
representation which can be found e.g. in \cite{GLP05}. Thus, there is enhanced supersymmetry in the near
horizon limit.

 The full solution interpolates between
 \begin{equation}
   \mbox{AdS}_3 \times X^7 \qquad \rightarrow \qquad \mathbb R^{1,1} \times c(X^7),
 \end{equation} 
 as expected for the $\alpha'$-corrected fundamental string. 
In the special case $X^7=S^7$ multi-instanton solutions and multi 5-branes have been constructed in \cite{Loginov05}, and it should be possible to generalize our solutions to include multi 5-branes in this case.

\subsection{Nearly K\"ahler}

The construction of a solution of heterotic string theory from a six-dimensional nearly K\"ahler manifold $X^6$ is almost identical to the nearly parallel $G_2$ case. We make the ansatz
\begin{align}
  M = \RR^{1,1} \times \RR \times X^6 \times S^1 \:,
\end{align}
with the metric
\begin{align}
  g = h^{-1}(-\dd t^2+\dd x^2) + e^{2f}(\dd\tau^2+g^6) + \dd y^2 \:,
\end{align}
where $y$ is a coordinate on $S^1$. By $e^a$ ($a=1,...,6$) we denote an orthonormal frame on $X^6$ and we set $e^7=\dd y$ and $e^8=\dd\tau$.

\subsubsection{Gravitino equation}

In the following, we will use the orthonormal frame
\begin{align}
 \sigma^0 &= h^{-1/2} \dd t \:, & 
 \sigma^a &= e^f e^a\:, & 
 \sigma^7 &= \dd y\:, &
 \sigma^8 &= e^f \dd\tau\:, & 
 \sigma^9 = h^{-1/2} \dd x \:.
\end{align}
We consider the following ansatz for $\nabla^-$:
\begin{align}
 \label{NK:connection1}
 \nabla^- = \nabla^P + s(\tau) e^aI_a + \zeta(\tau)\dd t\, I_8 + \xi(\tau)\dd x\, I_8 + \alpha(\tau) \dd\tau\, Z \:,
\end{align}
where $\nabla^P$ is the canonical SU$(3)$-connection on $X^6$ and $I_a$ are generators of the orthogonal complement of $\mathfrak{su}(3)$ in $\mathfrak{g}_2$.  As in the previous section, $I_8$ is one of the generators \eqref{R8-generators} corresponding to the $\tau$-direction and $Z$ is the $\mathfrak{so}(1,1)$-generator defined in \eqref{Zgenerator}.

The $T^0$, $T^8$ and $T^9$-components of the torsion are again given by \eqref{npG2:Ads3-torsion}, whereas $T^7=0$. The $T^a$-components were calculated in \cite{HN11}. Thus, requiring the torsion to be totally anti-symmetric results again in the first three equations in \eqref{G2-Hsol}. We obtain $T^\mu = \sigma^\mu\lrcorner H$ with
\begin{align}
  H = \dd h^{-1} \wedge \dd t \wedge \dd x - (\dot f -1)e^{2f}P \:,
\end{align}
where $P$ is the SU$(3)$-invariant 3-form on $X^6$. In order to make the $G_2 \ltimes \RR^8$ holonomy manifest, we perform a gauge transformation:
\begin{align}
 e^{-\frac 14 \log(h) Z}(\nabla^-)e^{\frac 14 \log(h) Z} 
  = \nabla^P + \dot f e^a I_a - \tfrac 12 e^{-f} \partial_\tau \log(h)(\dd t-\dd x)I_8 \:.
\end{align}
 There is again exactly one parallel spinor $\epsilon$.

\subsubsection{Dilatino equation}
The action of the 3-form $H$, as determined above, on the $G_2\ltimes \mathbb R^8$-invariant spinor is
\begin{align}
  H \cdot \epsilon = \left(-\partial_\tau \log(h) + 4(\dot f-1)\right) \dd\tau\cdot\epsilon \:.
\end{align}
Thus, the dilatino equation is solved by
\begin{align}
  \phi(\tau) = \phi_0 - \tfrac 12 \log(h) + 2(f-\tau) \:.
\end{align}

\subsubsection{Gaugino equation}
Analogously to the nearly parallel $G_2$ case we know that the connection
\begin{align}
 \label{NK:instanton connection}
  \nabla^P + \psi(\tau)e^aI_a
\end{align}
solves the $G_2$-instanton equation if and only if
\begin{align}
  \dot\psi = 2\psi(\psi-1) \:.
\end{align}
Thus, \eqref{NK:instanton connection} can be either the canonical connection $\nabla^P$ (for $\psi=0$), the Levi-Civita connection (for $\psi=1$) or the interpolating solution
\begin{align}
  \psi(\tau) = \left(1+e^{2(\tau-\tau_0)}\right)^{-1} \:.
\end{align}
We denote the curvature of \eqref{NK:instanton connection} by $\mathcal F(\psi)$ and set $\tilde R=\mathcal F(\psi_1)$ and $F=\mathcal F(\psi_2)$.

\subsubsection{Bianchi identity}
As the $\dd h^{-1} \wedge \dd t \wedge \dd x$-term in $H$ is obviously closed, the Bianchi identity is found, in close analogy to~\cite{HN11}, to be
\begin{align}
  (1-\dot f)e^{2f} = \frac{\alpha^\prime}{4} (\psi_1^2-\psi_1\dot\psi_1-\psi_2^2+\psi_2\dot\psi_2) \:.
\end{align}

\subsubsection{Field equations}

As in the nearly parallel $G_2$-case, the only equations of motion which are not trivially satisfied are the $B$-field equation and the $(tt)$-component of the Einstein equation, and these two coincide. For the $B$-field equation we calculate
\begin{align}
  \dd *e^{-2(\phi-\phi_0)}H = \dd\left[ \dot h e^{f+4\tau} \text{Vol}^7 \right] \:,
\end{align}
where Vol$^7$ is the volume form on $X^6 \times S^1$. Thus, the $B$-field equation reads
\begin{align}
  \partial_\tau(\dot h e^{f+4\tau}) = 0 \:.
\end{align}

\subsubsection{Solution}\label{sec:nk_sol}

We already know that the metric, the 3-form $H$ and the dilaton are given by
\begin{equation}\label{nk_10dsol}
\begin{aligned}
 g &= h^{-1} (-\dd t^2+\dd x^2) + e^{2f}(\dd\tau^2+g^6) + \dd y^2 \:,\\
 H &= \dd h^{-1}\wedge \dd t \wedge \dd x - (\dot f-1)e^{2f}P \:,\\
 \phi(\tau) &= \phi_0 -\tfrac 12 \log(h) + 2(f-\tau) \:.
\end{aligned}
\end{equation}
In the following, we will substitute $e^\tau =r$. In order to obtain the desired AdS$_3$-limit, we choose
\begin{align}
 \psi_1 = \frac{\rho^2}{\rho^2+r^2} \:, \qquad
 \psi_2 = 0 \:,
\end{align}
with some constant $\rho$. The Bianchi identity is solved by \cite{HN11}
\begin{equation}
 \begin{aligned}{}
      e^{2f} &= \lambda^2 r^2 + \frac {\alpha'}4 (\psi_1^2-\psi^2_2) \\ 
 &=\lambda^2 r^2+\frac {\alpha'}4\, \frac{\rho^4}{(\rho^2+ r^2)^2}   ,
 \end{aligned}
\end{equation}
with some constant $\lambda$, and from the $B$-field equation we obtain
\begin{equation}\label{nk_h_int}
  h(r) = a^{-2} + 5 Q_e \int_{\log(r)}^\infty \exp\Big(-f(\theta)-4\theta\Big) \dd\theta \:.
\end{equation}

In the limit $r\rightarrow\infty$ we obtain $\RR^{1,1}\times c(X^6) \times S^1$, with $c(X^6)$ being the Ricci-flat cone over $X^6$:
\begin{equation}\label{nk_sol_r_infty}
\begin{aligned}
 g &= a^{2}(-\dd t^2 + \dd x^2) +\lambda^2(\dd r^2+r^2g^6) + \dd y^2 \:, \\
 H &= 0 \:, \qquad
 e^{2(\phi-\phi_0)} = a^{2} \lambda^4 \:.
\end{aligned}
\end{equation}
In order to write the limit $r\rightarrow 0$ in a convenient way, we observe that $h^{-1}\rightarrow \frac{4}{5 Q_e} \left(\frac{\alpha^\prime}{4}\right)^{1/2} r^4 =: s^2$. Then the fields in this limit read
\begin{equation}\label{nk_sol_r_0}
\begin{aligned}
 g &= s^2(-\dd t^2+ \dd x^2) + \frac{\alpha^\prime}{16} \frac{\dd s^2}{s^2} + \frac{\alpha^\prime}{4} g^6 + \dd y^2 \:,\\
 H &= \dd(s^2)\wedge \dd t \wedge \dd x + \frac{\alpha^\prime}{4}P \:, \\
 e^{2(\phi-\phi_0)} &= \frac 4{5Q_e} \left(\frac{\alpha'}4\right)^{\textstyle{\frac 52}} \:.
\end{aligned}
\end{equation}
 The holonomy group of $\nabla^-$ reduces to that of $\nabla^P$, i.e. to SU(3). Table \ref{tab:stab} shows that there
are now four parallel spinors, and it turns out that two of them solve the dilatino equation. Again, in the near horizon
region we find twice as much supersymmetry as in the bulk. The full solution interpolates between
\begin{equation}
  \text{AdS}_3 \times X^6 \times S^1 \qquad \rightarrow \qquad \RR^{1,1} \times c(X^6) \times S^1 \:.
\end{equation}
 In the special case $X^6=S^6$ multi-instanton solutions and multi 5-branes are known \cite{Loginov05}, and we expect the solutions presented above to generalize to this case.

\subsection{Sasaki-Einstein}

In analogy with the previous two cases, we choose the space-time manifold to be of the form
\begin{equation}
 M = \mathbb{R}^{1,1} \times \mathbb{R} \times X^{2n+1} \times \mathbb{T}^{6-2n} \; ,
\end{equation}
where $X^{2n+1}$ is a Sasaki-Einstein manifold and $n=1,2,3$. The only r\^ole of the torus $\mathbb{T}^{6-2n}$ is to yield a 10-dimensional space-time as required for heterotic supergravity and, in particular, none of the fields depend on it. The metric is taken to be
\begin{equation}
 g  = h^{-1}(\tau) (-\dd t^2 +\dd x^2 ) + e^{2f(\tau)}(\dd \tau^2 + g_\ell^{2n+1}) + g_{\mathbb{T}^{6-2n}} \; ,
\end{equation}
where $t$ and $x$ are coordinates on $\mathbb{R}^{1,1}$, and $\tau$ parametrizes the remaining $\mathbb{R}$-factor. The Sasakian metric $g_\ell^{2n+1}$ on $X^{2n+1}$ in terms of a basis of 1-forms $(e^1, e^a)$, $a=2,\ldots,(2n+1)$, is given by
\begin{equation}
 g_\ell^{2n+1} = e^1 e^1 + e^{2\ell} \delta_{ab} e^a e^b \; ,
\end{equation}
which contains a deformation parameter $\ell$ that can be made $\tau$-dependent, i.e. $\ell=\ell(\tau)$. There exist two special values for $\ell$, namely $e^{2\ell}=1$ and $e^{2\ell}=2n/(n+1)$ \cite{HN11}. For reasons to be explained below, we are interested in solutions where the field $\ell(\tau)$ interpolates between these two values as $\tau\to\pm\infty$.

The SU$(n)$-invariant 3-form $P$ on $X^{2n+1}$ is normalized such that
\begin{equation}
 P = e^{123}+e^{145}+\dots +e^{1\,2n\,2n+1} \; .
\end{equation}

\subsubsection{Gravitino equation}

We make an ansatz for the connection $\nabla^-$ in the form  
\begin{equation}\label{SEnabla-ansatz}
 \nabla^- = \nabla^ P + t(\tau) e^1 I_1 + s(\tau) e^a I_a + \zeta(\tau) \dd t\, I_{2n+2} + \xi(\tau) \dd x\, I_{2n+2} + \alpha(\tau) \dd \tau \, Z,
\end{equation}
where $\nabla^P$ is the canonical SU$(n)$-connection on $X^{2n+1}$ and ($I_1$, $I_a$) are generators of the orthogonal
complement of $\mathfrak{su}(n)$ in $\mathfrak{su}(n+1)$. In addition, $I_{2n+2}$ is one of the
generators~\eqref{R8-generators} corresponding to the $\tau$-direction, and $Z$ is the
$\mathfrak{so}(1,1)$-generator~\eqref{Zgenerator}. The holonomy group of $\nabla^-$ is SU($n+1)\ltimes \mathbb R^8$,
and there are four parallel spinors if $n=1$ and two parallel spinors when $n=2,3$.

It is useful to introduce an orthonormal basis
\begin{equation}
\begin{aligned}
 \sigma^0 &= h^{-1/2} \dd t ,&\quad \sigma^{2n+3} &= h^{-1/2} \dd x ,&\quad \sigma^{i+(2n+3)} &= \dd y^i,\\
 \sigma^{2n+2} &= e^f\dd\tau ,&\quad \sigma^1 &= e^f e^1 , &\quad \sigma^a &= e^{f+\ell} e^a .
\end{aligned}
\end{equation}
Here, $y^i$ (with $i=1,\ldots,(6-2n)$) are coordinates on $\mathbb{T}^{6-2n}$. Using the Cartan structure equation $T^\mu = \dd\sigma^\mu + {}^-{\Gamma^\mu}_\nu \wedge \sigma^\nu$ we can calculate the torsion of $\nabla^-$; the $T^1$-, $T^a$-components from \cite{HN11} are unchanged, whereas $T^0$, $T^{2n+3}$ and $T^{2n+2}$ agree with their respective counterparts, $T^0$, $T^9$ and $T^8$, in~\eqref{npG2:Ads3-torsion}, because of the common form of the $\mathbb{R}^{1,1} \times \mathbb{R}$-part of the metric. In addition, one finds $T^{i+(2n+3)}=0$ for $i=1,\ldots,(6-2n)$. From $T^\mu = \sigma^\mu \lrcorner H$, we obtain the following conditions
\begin{equation}
\begin{aligned}
    \zeta = -\xi = e^{-f}&\partial_\tau h^{-1/2} ,\qquad    4 \alpha = -\partial_\tau \log(h),\qquad t=\dot f ,\qquad s=e^\ell (\dot{f} + \dot{\ell}) ,\\
    H&= \dd h^{-1} \wedge \dd t \wedge \dd x - \left( \frac{n+1}{n} (\dot{f}-1) + \dot{\ell} \right) e^{2(f+\ell)} P,
\end{aligned}
\end{equation}
where $\dot{(\ )} =\partial_\tau$. Moreover, we learn that
\begin{equation}
 \frac{n-1}{n} \dot{f} + \dot{\ell} = 2 e^{-2\ell} - \frac{n+1}{n} \; .
\end{equation}

\subsubsection{Dilatino equation}

The action of the 3-form $H$, as determined above, on an SU($n+1)\ltimes \mathbb R^{2n+2}$-invariant spinor $\epsilon$
is given by
\begin{equation}
 H \cdot \epsilon =\Big(- \partial_\tau \log(h) + (n+1)(\dot{f}-1) + n\dot{\ell} \Big) \dd\tau \cdot \epsilon \; .
\end{equation}
Hence, the dilatino equation $(\dd\phi - \frac 12 H)\cdot \epsilon=0$ is solved by
\begin{equation}\label{SE-phisol}
 \phi (\tau ) =\phi_0  -\sfrac 12 \log(h)   + \frac{n+1}{2} (f-\tau) + \frac{n}{2} \ell \; .
\end{equation}

\subsubsection{Gaugino equation}

The gaugino equation requires the gauge field to be a SU($n+1$)$\ltimes \mathbb R^{2n+2}$-instanton, and we also impose this condition on the connection $\tilde \nabla$. A connection of the form
\begin{equation}\label{SE-gaugefieldansatz}
 \nabla^P + \chi(\tau) e^1 I_1 + \psi(\tau) e^a I_a \; ,
\end{equation}
solves the SU($n+1$)-instanton equation if and only if $\chi$ and $\psi$ satisfy~\cite{HN11}
\begin{align}
 \dot\chi &= 2n e^{-2\ell} (\psi^2 - \chi) \; , \label{SE-chiinsteq} \\
 \dot\psi &= \frac{n+1}{n} \psi (\chi - 1) \; . \label{SE-psiinsteq}
\end{align}
There are two fixed points $(\psi,\chi)=(0,0)$ and $(1,1)$ which correspond to the canonical connection $\nabla^P$ and the Levi-Civita connection of the cone $c(X^{2n+1})$, respectively. Due to the non-linearity and the coupling to $\ell$, it is in general not possible to solve eqs.~\eqref{SE-chiinsteq}-\eqref{SE-psiinsteq} analytically. 

Denote the curvature form of \eqref{SE-gaugefieldansatz} by $\mathcal{F}(\psi,\chi)$. We put $\tilde R = \mathcal{F}(\psi_1,\chi_1)$, $F = \mathcal{F}(\psi_2,\chi_2)$ and later we shall choose $\psi_2 = \chi_2 = 0$.

\subsubsection{Bianchi identity}

Since the term $\dd h^{-1} \wedge \dd t\wedge \dd x$ in $H$ is closed, the Bianchi identity essentially reduces to the
same equation as in \cite{HN11}:
\begin{equation}\label{SE-Bianchi1storder}
 (\dot{f} + \dot{\ell} - e^{-2\ell}) e^{2(f+\ell)} = \frac{\alpha^\prime (n+1)}{8n} \left( \chi_2^2 - 2\chi_2 \psi_2^2 + 2\psi_2^2 - \chi_1^2 + 2\chi_1 \psi_1^2 - 2\psi_1^2 \right) .
\end{equation} 

\subsubsection{Field equations}

Again, the $B$-field equation coincides with the $(tt)$-component of the Einstein equation. We have
\begin{equation}
 \dd *e^{-2(\phi-\phi_0)} H = \dd\left[ \dot h\, \exp\Big( (n-1)f + n \ell + (n+1) \tau \Big) \text{Vol}^7 \right] \; ,
\end{equation} 
where Vol$^7$ denotes the volume form on $X^{2n+1} \times \mathbb{T}^{6-2n}$.
The $B$-field equation becomes
\begin{equation}\label{SE-Bfieldeq}
 \partial_\tau \Big[\dot h\, \exp\Big( (n-1)f + n \ell + (n+1) \tau \Big)\Big]=0.
\end{equation}

\subsubsection{Solution}\label{sec:se_sol}

We have arrived at the following form of the 10-dimensional fields
\begin{equation}\label{se_10dsol}
\begin{aligned}
      g  &= h^{-1}(\tau) (-\dd t^2 +\dd x^2 ) + e^{2f(\tau)}(\dd \tau^2 + e^1 e^1 + e^{2\ell(\tau)} \delta_{ab} e^a e^b) + g_{\mathbb{T}^{6-2n}}\; ,\\
      H&= \dd h^{-1} \wedge \dd t \wedge \dd x - \left( \frac{n+1}{n} (\dot{f}-1) + \dot{\ell} \right) e^{2(f+\ell)} P\; , \\
      \phi (\tau ) &=\phi_0  -\sfrac 12 \log(h)   + \frac{n+1}{2} (f-\tau) + \frac{n}{2} \ell \; ,
\end{aligned}
\end{equation}
which are determined in terms of
\begin{equation}\label{SE_PDEsystem}
\begin{aligned}
 &\frac{n-1}{n} \dot{f} + \dot{\ell} = 2 e^{-2\ell} - \frac{n+1}{n}, \\
 &\dot\chi_1 = 2n e^{-2\ell} (\psi_1^2 - \chi_1) , \qquad\quad
 \dot\psi_1 = \frac{n+1}{n} \psi_1 (\chi_1 - 1), \\
 &\dot\chi_2 = 2n e^{-2\ell} (\psi_2^2 - \chi_2) , \qquad\quad
 \dot\psi_2 = \frac{n+1}{n} \psi_2 (\chi_2 - 1), \\
 &(\dot{f} + \dot{\ell} - e^{-2\ell}) e^{2(f+\ell)} = \frac{\alpha^\prime (n+1)}{8n} \left( \chi_2^2 - 2\chi_2 \psi_2^2 + 2\psi_2^2 - \chi_1^2 + 2\chi_1 \psi_1^2 - 2\psi_1^2 \right), \\
 &\dot h = - 2n Q_e \exp\Big( -(n-1)f - n \ell - (n+1) \tau \Big) \; .
\end{aligned}
\end{equation}
This is a set of seven coupled non-linear ODEs for the seven unknown functions $f$, $\ell$, $h$, $\psi_1$, $\chi_1$, $\psi_2$, $\chi_2$. In general, the system of equations is sufficiently complicated such that analytic solutions are not attainable. A notable exception is the case $n=1$, i.e. 3D Sasaki-Einstein, which can be solved analytically and will be discussed below. 

\paragraph{3D Sasaki-Einstein ($n=1$).}
The only simply connected 3-dimensional Sasaki-Einstein manifold is the 3-sphere $S^3=\mbox{SU}(2)$, and this case
was considered in \cite{LLOP99}. The Bianchi identity assumes a slightly more general form than in the higher
dimensional examples. Upon setting $\ell=0$ the gravitino equation yields the following result for the 3-form:
 \begin{equation}
   H = \dd h^{-1}\wedge \dd t\wedge dx -2(\dot f-1) e^{2f} P,
 \end{equation}  
 with $P=\mbox{Vol}_{S^3}$. The right hand side of the Bianchi identity is determined as
 \begin{equation}
  \mbox{tr} \Big( \tilde R\wedge \tilde R - F\wedge F \Big )= 12  \dd\big(6\psi^2 -4\psi^3 \big)P,
 \end{equation} 
 where we set $F=\mathcal F(0,0)$ and $\tilde R = \mathcal F(\psi,\psi)$. Since $P$ is closed, the Bianchi identity
 implies
 \begin{equation}
   (\dot f-1) e^{2f}  = \frac {\alpha'}4(2\psi^3-3\psi^2)-Q_m ,
 \end{equation} 
 where $Q_m$ is an integration constant, to be identified with an NS5-brane charge. Since the canonical 3-form
$P$ is not closed for the other geometries we consider, it is not possible to add the $Q_m$-term to the Bianchi identity
in these cases. The system of equations \eqref{SE_PDEsystem} reduces to
\begin{equation}
 \dot\psi = 2 \psi (\psi - 1), \qquad\;\;
 (\dot{f} - 1) e^{2f} = \frac{\alpha^\prime}{4} \psi^2 ( 2\psi - 3 )-Q_m, \qquad\;\;
 \dot h = - 2 Q_e e^{-2\tau} \; ,
\end{equation}
which is solved, in terms of a radial coordinate $r=e^\tau$, by
\begin{equation}
\begin{aligned}
 h &= a^{-2} + \frac{Q_e}{r^2} ,\qquad  \psi =  \frac{\rho^2}{\rho^2 + r^2}\\
 e^{2f} &= \lambda^2 r^2 + \frac {\alpha'}4 \psi^2 +Q_m =\lambda^2 r^2 + \frac {\alpha'}4 \frac{\rho^4}{(\rho^2 +
r^2)^2}+Q_m ,
\end{aligned}
\end{equation} 
with constants $a,\lambda,\rho,Q_e,Q_m \in \mathbb R$. The full 10-dimensional solution is then of the form
\begin{equation}\label{SE3_sol_10dfields}
\begin{aligned}
 g &= \frac {r^2}{r^2/a^2 +Q_e}(-\dd t^2+\dd x^2) +\Big(\lambda^2 + \frac{\alpha'}4 \frac{\psi^2}{r^2} + \frac{Q_m}{r^2}
\Big) \big(\dd r^2 +r^2 g_{S^3}\big) + g_{\mathbb{T}^4}\; ,\\
 H &= \frac{2Q_er}{(r^2/a ^ 2+Q_e) ^2} \dd r \wedge \dd t\wedge \dd x +\Big(2Q_m- \frac{\alpha'}2 \psi^2 ( 2\psi - 3
)\Big) \mbox{Vol}_{S^3}\; , \\
 e^{2(\phi-\phi_0)} &= \frac {1}{r^2/a^2 + Q_e} \Big(\lambda^2 r^2+ \frac{\alpha'}4{\psi^2} +Q_m\Big).
\end{aligned}
\end{equation}
This is the `gauge dyonic string' of \cite{Duff:gaugedyonicstring}.
In the limit $r\rightarrow 0$ the above fields become
\begin{equation}\label{S3nhlimit}
\begin{aligned}
 g & = \frac {r^2}{Q_e} (-\dd t^2+\dd x^2) + \Big(\frac {\alpha'}4+Q_m\Big)  \frac{\dd r^2}{r^2} +  \Big(\frac
{\alpha'}4+Q_m\Big) g_{S^3} + g_{\mathbb{T}^4}\; ,\\
 H&= \frac {2r}{Q_e} \dd r\wedge \dd t\wedge \dd x + 2\Big(\frac {\alpha'}4+Q_m\Big) \,\mbox{Vol}_{S^3} \;, \\
 e^{2(\phi-\phi_0)} &= \frac 1{Q_e} \Big(\frac {\alpha'}4+Q_m\Big) ,
\end{aligned}  
\end{equation}
describing an AdS$_3\times S^3\times \mathbb T^4$ geometry. The holonomy group of $\nabla^-$ is trivial, and half of all parallel spinors
satisfy the dilatino equation, giving rise to eight preserved supersymmetries.
The other limit $r\rightarrow \infty$ is, at least for $\lambda\neq 0$:
\begin{equation}
\begin{aligned}
 g &= a^2(-\dd t^2+\dd x^2)  +\lambda^2 (\dd r^2 +r^2 g_{S^3}) + g_{\mathbb{T}^4}\;,\\
 H&= 0\;, \qquad e^{2(\phi-\phi_0)} = a^2\lambda^2,
\end{aligned}
\end{equation}
which is the Ricci-flat solution $\mathbb R^{1,1} \times c(S^3)$, where $c(S^3) = \mathbb R^4\setminus\{0\}$ denotes the
cone over $S^3$. Thus, the solution interpolates between
\begin{equation}
 \mbox{AdS}_3 \times S^3 \times \mathbb T^4  \qquad \rightarrow \qquad \mathbb R^{1,1} \times \mathbb R^4 \times \mathbb T^4 \; .
\end{equation}
 In the limit $\rho\rightarrow \infty$, or $\psi=1$ we obtain a solution without
$\alpha'$-corrections (except for the singularity at $r=0$), hence vanishing field strength of the gauge fields, and a new $Q_m'$,
 \begin{equation}
   Q_m '  = Q_m  + \frac {\alpha'}4.
 \end{equation} 
 This justifies our interpretation of $Q_m$ as an NS5-brane charge. Based on multi-instanton gauge fields it is also possible to
find gauge multi-brane solutions \cite{LLOP99}. An interesting special case of the above solution occurs for $a^{-1} = \lambda=0$ \cite{Gava10}. Then the solution interpolates between AdS$_3\times S^3$ in both limits $r\rightarrow 0$ and $r\rightarrow \infty$, but with different radii. This is interpreted as a renormalization group flow of the dual CFT in \cite{Gava10}.

\begin{figure}[t]
\includegraphics[width=\textwidth, clip]{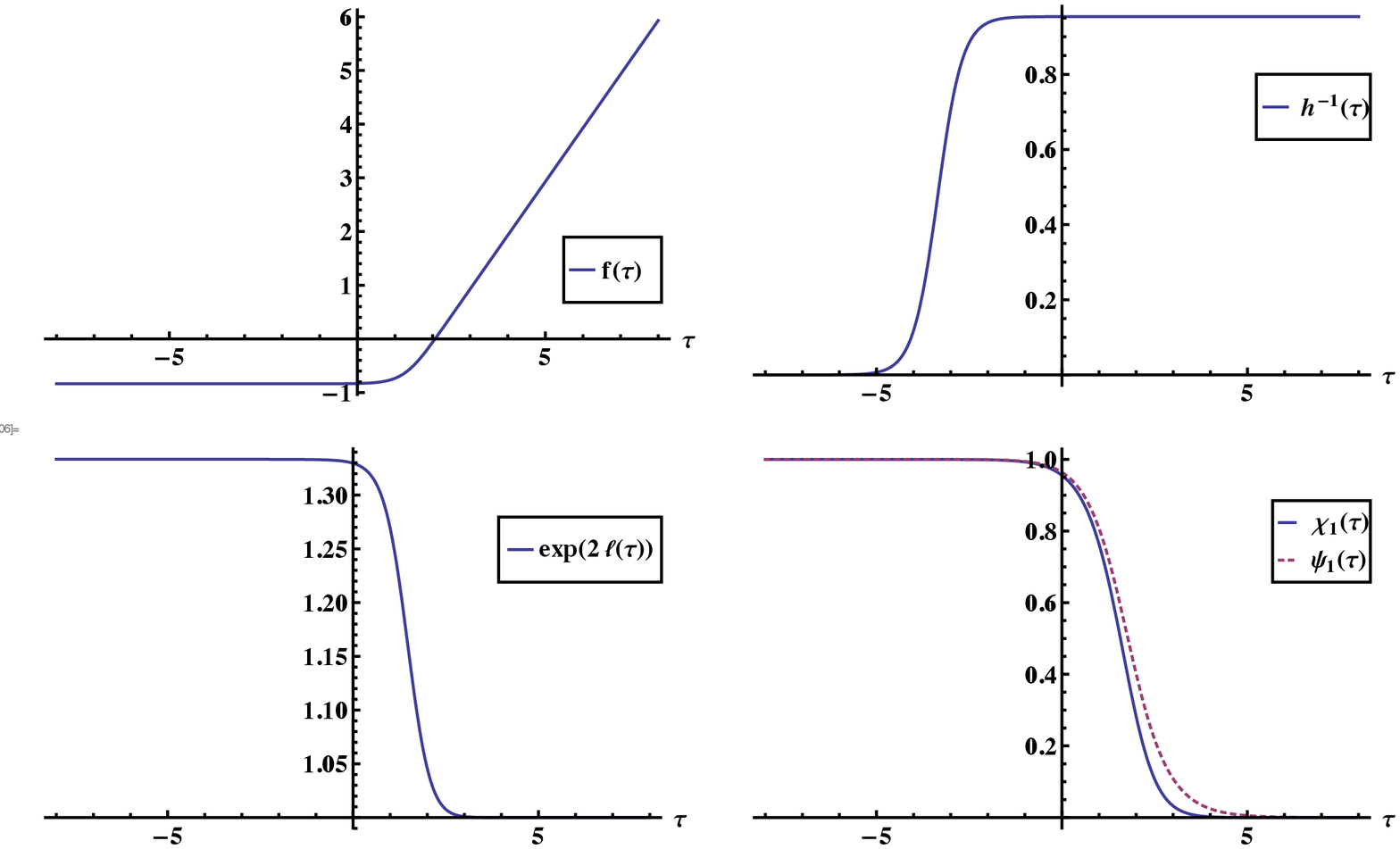}
\caption{Numerical solution with $\psi_2 = \chi_2 = 0$ for the 5D Sasaki-Einstein case. The solution interpolates between $\mbox{AdS}_3\times X^5 \times\mathbb{T}^2 \rightarrow \mathbb{R}^{1,1} \times c(X^5) \times\mathbb{T}^2$ as $\tau$ changes from $-\infty$ to $+\infty$.}
\label{fig:se5sol}
\end{figure}

\begin{figure}[t]
\includegraphics[width=\textwidth, clip]{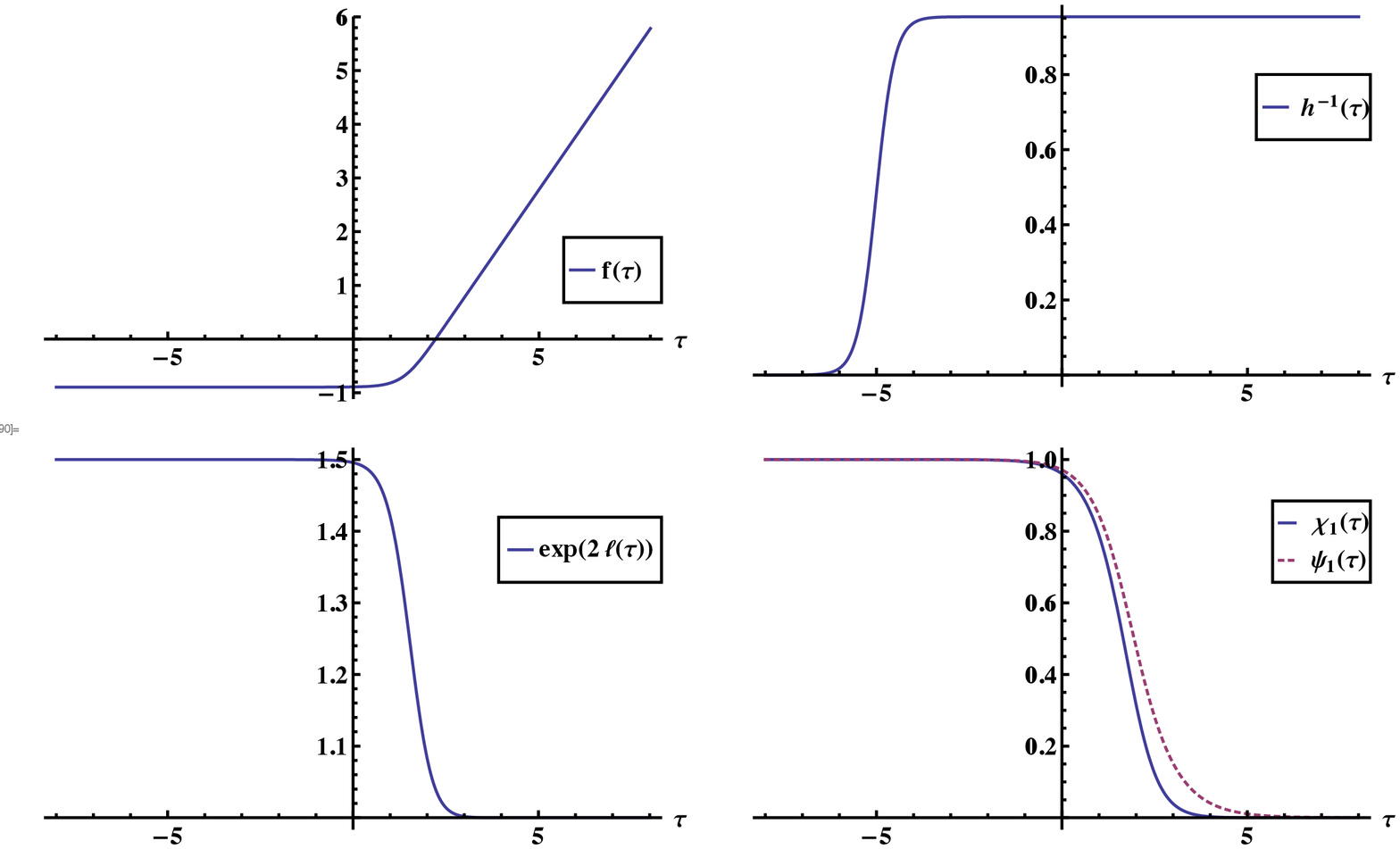}
\caption{Numerical solution with $\psi_2 = \chi_2 = 0$ for the 7D Sasaki-Einstein case. The solution interpolates between $\mbox{AdS}_3\times X^7 \rightarrow \mathbb{R}^{1,1} \times c(X^7)$ as $\tau$ changes from $-\infty$ to $+\infty$.}
\label{fig:se7sol}
\end{figure}

\paragraph{5D and 7D Sasaki-Einstein ($n=2,3$).}
For $n>1$, the equations do not decouple and hence we need to resort to numerical solutions. We set $\alpha'=1$, for
convenience, and choose $\psi_2 = \chi_2 = 0$. With an appropriate choice of boundary values we indeed find solutions
with the desired asymptotic behaviour, both for $n=2$ and $n=3$. Exemplary numerical solutions for $n=2$ and $n=3$ are
presented in Figures~\ref{fig:se5sol} and~\ref{fig:se7sol}, respectively.

As $\tau\to-\infty$, which corresponds to $r\to 0$ for the radial coordinate $r=e^\tau$, the one-dimensional fields display the following limiting behaviour
\begin{equation}
\begin{aligned}
 e^{2f} &\to \frac{\alpha' (n+1)}{8n} \; , \quad\qquad h^{-1} \to Q_e^{-1} \sqrt{\frac{n+1}{2n}} \left(\frac{\alpha'}{4}\right)^{(n-1)/2} r^{n+1} \; , \\
 e^{2\ell} &\to \frac{2n}{n+1} \; , \qquad \psi_1 \to 1 \; , \qquad \chi_1 \to 1 \; .
\end{aligned}
\end{equation}
With $s^2:=Q_e^{-1} \sqrt{\frac{n+1}{2n}} \left(\frac{\alpha'}{4}\right)^{(n-1)/2} r^{n+1}$, the 10-dimensional fields thus become
\begin{equation}\label{se_sol_r_0}
\begin{aligned}
 g  &= s^2 (-\dd t^2 +\dd x^2 ) + \frac{\alpha'}{2n(n+1)} \frac{\dd s^2}{s^2} + \frac{\alpha' (n+1)}{8n} g_\ell^{2n+1} + g_{\mathbb{T}^{6-2n}}\;,\\
 H&= \dd (s^2) \wedge \dd t \wedge \dd x + \frac{\alpha' (n+1)}{4n} P \;,\\
 e^{2(\phi-\phi_0)} &= \frac{n+1}{2n Q_e} \left(\frac{\alpha'}{4}\right)^n  \; ,
\end{aligned}
\end{equation}
which describes the direct product $\mbox{AdS}_3\times X^{2n+1} \times \mathbb{T}^{6-2n}$. The holonomy group of
$\nabla^-$ reduces to SU($n$), which stabilizes eight spinors for $n=2$ and four for $n=3$. In each case only half of
all parallel spinors satisfy the dilatino equation, so that there remain four Killing spinors for $n=2$ and two for
$n=3$.

In the limit $r\to\infty$, the one-dimensional fields approach the following values
\begin{equation}
\begin{aligned}
 e^{2f} &\to \lambda^2 r^2 \; , \qquad h^{-1} \to a^2 \; , \\
 e^{2\ell} &\to 1 \; , \qquad \psi_1 \to 0 \; , \qquad \chi_1 \to 0 \; ,
\end{aligned}
\end{equation}
with constants $\lambda,a\in\mathbb{R}$. The corresponding 10-dimensional fields take the form
\begin{equation}\label{se_sol_r_infty}
\begin{aligned}
 g  &= a^2 (-\dd t^2 +\dd x^2 ) + \lambda^2 (\dd r^2 + r^2 g_\ell^{2n+1}) + g_{\mathbb{T}^{6-2n}}\;,\\
 H&= 0\; , \qquad e^{2(\phi-\phi_0)} = a^2\, \lambda^{n+1} \; .
\end{aligned}
\end{equation}
Up to a coordinate rescaling, this describes the Ricci-flat cone solution $\mathbb{R}^{1,1} \times c(X^{2n+1}) \times \mathbb{T}^{6-2n}$.

In conclusion, the numerical solutions presented in Figures~\ref{fig:se5sol} and~\ref{fig:se7sol} interpolate between
\begin{equation}
 \mbox{AdS}_3\times X^{2n+1} \times\mathbb{T}^{6-2n} \qquad \rightarrow \qquad \mathbb{R}^{1,1} \times c(X^{2n+1}) \times\mathbb{T}^{6-2n} \; .
\end{equation}

\subsection{3-Sasakian}\label{sec:3S}

Let $X^7$ be a 7-dimensional 3-Sasakian manifold. We make the same ansatz for the space-time manifold and its metric as in the previous section (for $n=3$), namely
\begin{equation}
 M = \RR^{1,1} \times \RR \times X^7 \:, \quad
 g = h^{-1}(\tau)(-\dd t^2+\dd x^2) + e^{2f(\tau)}(\dd\tau^2+g^7_\ell) \:,
\end{equation}
where $t$ and $x$ are coordinates on $\RR^{1,1}$ and $\tau$ parametrizes the $\RR$-factor. Furthermore, by ($e^\alpha$, $e^a$), $\alpha=1,2,3$, $a=4,\ldots,7$, we denote an orthonormal basis of one-forms on $X^7$ and define $e^8:=\dd\tau$. As in the Sasaki-Einstein case, the metric $g^7_\ell$ depends on a deformation parameter $\ell$, which will be promoted to a $\tau$-dependent function. \vspace{\baselineskip}

Associated to the Sasaki-Einstein structures on $X^7$ there are three one- and three two-forms $\eta^\alpha$ and $\omega^\alpha$. We choose the frame ($e^\alpha$, $e^a$) such that they are given by
\begin{equation}
 \begin{aligned}
  \eta^1&= e^1\:, &\quad \omega^1 &= e^{45}+e^{67} \:,\\
  \eta^2&= e^2\:, &\quad \omega^2 &= e^{46}-e^{57} \:,\\
  \eta^3&= e^3\:, &\quad \omega^3 &= e^{47}+e^{56} \:.
 \end{aligned}
\end{equation}
In this frame, the metric on $X^7$ is
\begin{align}
 g_\ell^7 = \delta_{\alpha\beta}e^\alpha e^\beta + e^{2\ell(\tau)} \delta_{ab} e^a e^b \: ,
\end{align}
and the Sp$(1)$-invariant 3-form $P$ is normalized such that
\begin{equation}
 P = \frac13\eta^{123} + \frac13\eta^\alpha\wedge\omega^\alpha \; .
\end{equation}

\subsubsection{Gravitino equation}

In the following, we will use the orthonormal frame
\begin{equation}
\begin{aligned}{}
 \sigma^0 &= h^{-1/2} \dd t \:, \:\: & 
 \sigma^\alpha &= e^f e^\alpha \:, & 
 \sigma^a &= e^{f+\ell} e^a\:, \\
 \sigma^8 &= e^f \dd\tau\:, & 
 \sigma^9 &= h^{-1/2} \dd x \:.
\end{aligned}
\end{equation}
We make an ansatz for the $\nabla^-$ connection of the form
\begin{align}
 \nabla^- = \nabla^P + t(\tau)e^\alpha I_\alpha + s(\tau)e^a I_a + \zeta(\tau) \dd t\, I_8 + \xi(\tau) \dd x\, I_8 + \alpha(\tau) \dd\tau \, Z \:,
\end{align}
where $\nabla^P$ is the canonical Sp$(1)$-connection on $X^7$ and ($I_\alpha$, $I_a$) are generators of the orthogonal
complement of $\mathfrak{sp}(1)$ in $\mathfrak{sp}(2)$. The holonomy group is Sp($2)\ltimes \mathbb R^8$, giving rise
to three parallel spinors.
The $T^\alpha$-, $T^a$-components of the torsion were already
calculated in \cite{HN11} and the $T^0$-, $T^8$- and $T^9$-components are given again by \eqref{npG2:Ads3-torsion}.
Thus, requiring the torsion to be totally antisymmetric, i.e. $T^\mu = \sigma^\mu \lrcorner H$ for some 3-form $H$,
results in the conditions
\begin{align}
 t &= \dot f \:, &
 s &= e^\ell(\dot f+\dot \ell) \:, &
 \zeta = -\xi = e^{-f} \partial_\tau h^{-1/2} \:, \\
 4\alpha &= -\partial_\tau\log(h)\:, &
 \dot f + \dot \ell &= 2 e^{-2\ell} -1,
\end{align}
and the 3-form $H$ is given by
\begin{align}
 H &= \dd h^{-1} \wedge \dd t \wedge \dd x + H^{(8)} \:,\\
 H^{(8)} &= -2e^{2f}(\dot f-1)\eta^{123} -e^{2(f+\ell)}(\dot f+\dot \ell-1)\eta^\alpha\wedge\omega^\alpha \:.
\end{align}
To show that the connection $\nabla^-$ has Sp$(2)\ltimes\RR^8$-holonomy, and hence two parallel spinors, we perform a gauge transformation
\begin{multline}
  e^{-\frac 14 \log(h) Z}(\nabla^-)e^{\frac 14 \log(h) Z} 
  = \nabla^P + \dot f e^\alpha I_\alpha + e^\ell(\dot f+\dot \ell) e^a I_a + e^{-f} \partial_\tau\log(h) (\dd t-\dd x) I_8 \:.
\end{multline}

\subsubsection{Dilatino equation}
For the action of the 3-form $H$, as determined above, on an Sp$(2)\ltimes \mathbb R^8$-invariant spinor $\epsilon$ we
obtain
\begin{align}
 H\cdot\epsilon = \left(- \partial_\tau\log(h) + 4(\dot f-1) + 2\dot \ell \right) \dd\tau\cdot\epsilon \:.
\end{align}
Thus, the dilatino equation, $(d\phi-\frac 12 H)\cdot\epsilon =0$, is solved by
\begin{align}
 \phi(\tau) = \phi_0 -\tfrac12 \log(h) + 2(f-\tau) + \ell \:.
\end{align}

\subsubsection{Gaugino equation}
For solving the gaugino equation we revert to the Sp$(2)$-instanton solution constructed in \cite{HN11}. We know that the connection
\begin{align}
\label{3S:instanton connection}
 \nabla^P + \chi(\tau) e^\alpha I_\alpha + \psi(\tau)e^a I_a \:,
\end{align}
gives an instanton if $\chi$ and $\psi$ satisfy
\begin{align}
 0 &= \chi - \psi^2 \:,\\
 \dot \chi &= 2\chi(\chi-1) \:,\\
 \dot \psi &= \psi(\chi-1) \:.
\end{align}
These equations admit the constant solutions $\chi=\psi=0$, $\chi=\psi=1$ and a solution interpolating between the two, namely
\begin{align}
 \chi &= \psi^2 = \left( 1+e^{2(\tau-\tau_0)} \right)^{-1} \:.
\end{align}
We denote the curvature of \eqref{3S:instanton connection} by $\mathcal F(\chi,\psi)$ and set $\tilde R=\mathcal F(\chi_1,\psi_1)$ and $F=\mathcal F(\chi_2,\psi_2)$.

\subsubsection{Bianchi identity}
The Bianchi identity is equivalent to the two equations~\cite{HN11}
\begin{align}
 e^{2f}(\dot f - 1) &= \frac{\alpha^\prime}{8} \left( 2\chi_1^3 -3\chi_1^2 - 2\chi_2^3 + 3\chi_2^2  \right) \:,\\
 e^{2(f+\ell)}(\dot f + \dot \ell - 1) &= \frac{\alpha^\prime}{4} \left( \chi_1^2 -2\chi_1 - \chi_2^2 + 2\chi_2 \right) \:.
\end{align}
Furthermore, these equations are solved by
\begin{align}
 e^{2f} &= \lambda^2 e^{2\tau} + \frac{\alpha^\prime}{8} \left( \chi_1^2 - \chi_2^2 \right) \:,\\
 e^{2\ell} &= 2\,\frac{4\lambda^ 2 e^{2\tau}+\alpha^\prime(\chi_1-\chi_2)}{8\lambda^2 e^{2\tau}+\alpha^\prime(\chi_1^2-\chi_2^2)}\:.
\end{align}

\subsubsection{Equations of motion}

As in the previous cases, the only equations of motion which are not trivially satisfied are the $B$-field equation and the $(tt)$-component of the Einstein equation. For solving the $B$-field equation we calculate
\begin{align}
 \label{3S:*H}
 \dd *e^{-2(\phi-\phi_0)}H = \dd \left[ \dot h e^{2f+4\tau+2\ell} \text{Vol}^7 \right] \:,
\end{align}
where Vol$^7$ denotes the volume form on $X^7$. The $B$-field equation becomes
\begin{align}
\label{3S:B-field equation}
 \partial_\tau\left[ \dot h e^{2f+4\tau+2\ell} \right] = 0 \:.
\end{align}
It turns out that the $(tt)$-component of the Einstein equation coincides with \eqref{3S:B-field equation}.

\subsubsection{Solution}\label{sec:3s_sol}

We already know that the metric, the 3-form $H$ and the dilaton are given by
\begin{equation}\label{3s_10dsol}
\begin{aligned}
 g &= h^{-1}(-\dd t^2+\dd x^2) + e^{2f}(\dd\tau^2+g_\ell^7) \:,\\
 H &= \dd h^{-1}\wedge \dd t\wedge \dd x -2e^{2f}(\dot f-1)\eta^{123} -e^{2(f+\ell)}(\dot f+\dot \ell-1)\eta^\alpha\wedge\omega^\alpha \: , \\
 \phi &= \phi_0 - \frac 12 \log(h) + 2 (f-\tau) + \ell \: .
\end{aligned}
\end{equation}
The $B$-field equation can be rewritten as
\begin{equation}
  \dot h = - 6 Q_e e^{-2f-4\tau-2\ell} \:.
\end{equation}
Thus, a solution for $h$ can be calculated from $f$ and $\ell$ by
\begin{equation}\label{3s_h_int}
  h (r) = a^{-2} + 6 Q_e \int_{\log(r)}^\infty \exp\Big(-2f(\theta)-4\theta-2\ell(\theta) \Big) \dd\theta \:,
\end{equation}
with a constant $a$.

In order to obtain a solution with an AdS$_3$-limit, we choose
\begin{equation}
 \chi_1 = \psi_1^2 = \frac{\rho^2}{\rho^2+r^2} \:,\qquad
 \chi_2 = \psi_2^2 = 0 \:,
\end{equation}
with $r:=e^\tau$, as before, and some constant $\rho$. Therewith, the Bianchi identity yields
\begin{align}
 e^{2f} &= \lambda^2r^2 + \frac{\alpha^\prime}{8} \frac{\rho^4}{(\rho^2+r^2)^2} \; , \\
 e^{2\ell} &= 2(\rho^2+r^2)\,\frac{4\lambda^ 2 r^2(\rho^2+r^2)+\alpha^\prime\rho^2}{8\lambda^2 r^2(\rho^2+r^2)^2+\alpha^\prime\rho^4} \; ,
\end{align}
and the integral expression~\eqref{3s_h_int} for $h(r)$ can be explicitly computed
\begin{equation}\label{3s_h_sol}
\begin{split}
  h &= a^{-2} + 6Q_e \left[ 
     \frac{1}{\alpha^\prime r^4}
    +\frac{2 (\alpha'-4\rho^2\lambda^2)}{\alpha^{\prime 2}\rho^2 r^2}
    -\frac{8\lambda^2(\alpha'-2\rho^2\lambda^2)}{\alpha^{\prime 3}\rho^2} \log\left(1+\frac{\rho^2}{r^2}+\frac{\alpha^\prime\rho^2}{4\lambda^2r^4}\right) \right. \\
  &\qquad\left.
    +\frac{4\lambda^2(8\rho^2\lambda^2\alpha'-\alpha'^2-8\rho^4\lambda^4)}{\alpha^{\prime 3}\rho^2}
    \begin{cases}
      \frac{1}{\rho\lambda\sqrt{\alpha^\prime-\rho^2\lambda^2}}\left(\frac{\pi}{2}-\arctan\left(\frac{\lambda(2r^2+\rho^2)}{\rho\sqrt{\alpha^\prime-\rho^2\lambda^2}}\right)\right)
      &\text{for}\quad \rho^2\lambda^2<\alpha^\prime \\
      \frac{1}{2\lambda^2r^2+\alpha^\prime}
      &\text{for}\quad \rho^2\lambda^2=\alpha^\prime \\
      \frac{1}{2\rho\lambda\sqrt{\rho^2\lambda^2-\alpha^\prime}}
      \log\left(\frac{2 \lambda r^2+\rho^2\lambda+\rho\sqrt{\rho^2\lambda^2-\alpha^\prime}}{2 \lambda r^2+\rho^2\lambda-\rho\sqrt{\rho^2\lambda^2-\alpha^\prime}} \right)
      &\text{for}\quad \rho^2\lambda^2>\alpha^\prime
    \end{cases}
  \right]
\end{split}
\end{equation}

\paragraph{Limit $r\rightarrow\infty$.} In this limit we obtain
\begin{equation}\label{3s_sol_r_infty}
\begin{aligned}
 g &= a^{2}(-\dd t^2+\dd x^2) + \lambda^2(\dd r^2+r^2g_\ell^7) \:, \\
 e^{2\ell} &= 1 \:, \\
 H &= 0 \:, \qquad
 e^{2(\phi-\phi_0)} = a^{2} \lambda^4 \:,
\end{aligned}
\end{equation}
which is $\RR^{1,1}\times c(X^7)$, where $c(X^7)$ is the Ricci-flat cone over $X^7$.

\paragraph{Limit $r\rightarrow 0$.}
In this limit we obtain an AdS$_3\times X^7$ geometry with
\begin{equation}\label{3s_sol_r_0}
\begin{aligned}
 g &= s^2(-\dd t^2+\dd x^2) + \frac{\alpha^\prime}{32}\frac{\dd s^2}{s^2} + \frac{\alpha^\prime}{8}g^7_\ell \:, \\
 s^2 &:= \frac{\alpha^\prime}{6 Q_e}r^4 \:, \qquad
 e^{2\ell} = 2 \:,\\
 H &= \dd(s^2)\wedge \dd t\wedge \dd x + \frac{3 \alpha^\prime}{4} P \:, \\
 e^{2(\phi-\phi_0)} &= \frac{1}{3 Q_e} \left(\frac{\alpha'}{4}\right)^3 \:.
\end{aligned}
\end{equation}
The holonomy group of $\nabla^-$ reduces to Sp(1), allowing for eight parallel spinors according to Table
\ref{tab:stab}. Of those, six also satisfy the dilatino equation.
We again obtain a solution interpolating between
\begin{equation}
  \text{AdS}_3 \times X^7 \qquad \longrightarrow \qquad \RR^{1,1} \times c(X^7),
\end{equation}
 with enhanced supersymmetry in the near horizon region.

\subsection{Fundamental strings with NS5-branes}\label{ssec:F1+NS5}
 The limit $\rho\rightarrow \infty$ or $\psi_1 \rightarrow 1$ eliminates the gauge anti-5-brane, and we are left with an NS1+NS5-brane system, or a fundamental string with an NS5-brane. The NS5-brane is wrapped on a calibrated cycle of dimension $k-3$, or a collection of those, in the cone $c(X^k)$. The case $X=S^3$ of an unwrapped NS5-brane has been studied for instance in \cite{LLOP99}. The gauge
connections $\tilde \nabla$ and $\nabla^A$ globally coincide with the Levi-Civita connection on the cone $c(X)$ and the
canonical connection on $X$, respectively, whereas the limiting behaviour of the other fields as $r\rightarrow
0,\infty$ remains unchanged.
 Note one minor difference between the cases $X=S^3$ and $\dim X>3$. In the former case we have vanishing curvature of
both the Levi-Civita connection on the cone and the canonical connection on $X$, hence the
$\alpha'$-corrections to the equations vanish and $H$ is closed, except for a $\delta$-function singularity at the origin. In higher dimensions there are non-vanishing $\alpha'$-corrections everywhere, and $H \propto P$ is not closed. Furthermore, our construction only yields solutions with one unit of brane
charge in the higher-dimensional setting, whereas for $X=S^3$ multi-brane solutions can be easily written down.

\paragraph{Nearly parallel $G_2$.}
Taking $\rho\to\infty$ in the solution found in Section~\ref{sec:g2_sol}, we obtain
\begin{equation}\label{g2_ns5_sol}
 \psi_1 = 1 \; ,\qquad e^{2f} = \lambda^2 r^2 + \frac{\alpha'}{4} \; .
\end{equation}
In this limit, we may also solve eq.~\eqref{g2_h_int} for $h$ explicitly
\begin{equation}
 h(r) = a^{-2} + \frac{18 Q_e}{7 \alpha'^3} \left[
    (2\alpha' + 8 r^2 \lambda^2)^{1/3} (\alpha'^2 - 6 r^2 \alpha' \lambda^2 + 72 r^4 \lambda^4) r^{-14/3} -144 \lambda^{14/3} \right] \; .
\end{equation}
The full 10-dimensional solution can be obtained straightforwardly by plugging the two expressions above into~\eqref{g2_10dsol}. The limiting solutions for $r\to 0,\infty$ coincide with the ones in Section~\ref{sec:g2_sol} given by~\eqref{g2_sol_r_infty}-\eqref{g2_sol_r_0}.

\paragraph{Nearly K\"ahler.}
The values for $\psi_1$ and $e^{2f}$ are the same as in eq.~\eqref{g2_ns5_sol}. To determine $h$, we solve eq.~\eqref{nk_h_int} and find
\begin{multline}
 h(r) = a^{-2} + \frac{5 Q_e}{2 \alpha'^{5/2}} \left[ \sqrt{\alpha'} \left(\frac{\alpha'}{r^4} - \frac{6 \lambda^2}{r^2} \right) \sqrt{\alpha' + 4 r^2 \lambda^2} \vphantom{\frac{1}{2}} \right.\\\left. - 
     12 \lambda^4 \log \left(\frac{2 r^2 \lambda^2}{\alpha' + 2 r^2 \lambda^2 + \sqrt{\alpha'} \sqrt{\alpha' + 4 r^2 \lambda^2}} \right) \right] \; .
\end{multline}
The full 10-dimensional solution is obtained by plugging the above expression together with~\eqref{g2_ns5_sol} into~\eqref{nk_10dsol}. As before, the limiting solutions for $r\to 0,\infty$ coincide with the ones in Section~\ref{sec:nk_sol} given by~\eqref{nk_sol_r_infty}-\eqref{nk_sol_r_0}.

\paragraph{Sasaki-Einstein.}
For $n=1$, the limit $\rho\to\infty$ is a special case of solution~\eqref{SE3_sol_10dfields}. Hence, the $r\to 0,\infty$ limits and the interpolating behaviour $\mbox{AdS}_3 \times S^3 \times \mathbb T^4  \rightarrow \mathbb R^{1,1} \times \mathbb R^4 \times \mathbb T^4$ remain unchanged.

For $n=2,3$, the limit $\rho\to\infty$ corresponds to setting $\psi_1 = \chi_1 = 1$ and $\psi_2 = \chi_2 = 0$. Eqs.~\eqref{SE_PDEsystem} then reduce to 
\begin{equation}
\begin{aligned}
 &\frac{n-1}{n} \dot{f} + \dot{\ell} = 2 e^{-2\ell} - \frac{n+1}{n}, \\
 &(\dot{f} + \dot{\ell} - e^{-2\ell}) e^{2(f+\ell)} = - \frac{\alpha^\prime (n+1)}{8n}, \\
 &\dot h = - 2n Q_e \exp\Big( -(n-1)f - n \ell - (n+1) \tau \Big) .
\end{aligned}
\end{equation}
Although considerably simpler than~\eqref{SE_PDEsystem}, this system of equations still appears to not be solvable analytically. However, it is possible to find numerical solutions that have the same limiting behaviour as the solutions found in Section~\ref{sec:se_sol}, i.e. $\mbox{AdS}_3\times X^{2n+1} \times\mathbb{T}^{6-2n} \rightarrow \mathbb{R}^{1,1} \times c(X^{2n+1}) \times\mathbb{T}^{6-2n}$ with the 10-dimensional fields approaching~\eqref{se_sol_r_0} and~\eqref{se_sol_r_infty}. The graphs for $f(\tau)$, $h^{-1}(\tau)$ and $e^{2\ell(\tau)}$ closely resemble those of Figures~\ref{fig:se5sol} and~\ref{fig:se7sol} and are thus omitted.

\paragraph{3-Sasakian.}
If we take the limit $\rho\to\infty$ of the general solution obtained in Section~\ref{sec:3s_sol}, we find
\begin{equation}
 e^{2f} = \lambda^2 r^2 + \frac{\alpha'}{8} \; , \qquad
 e^{2\ell} = \frac{8\lambda^2 r^2 + 2\alpha'}{8\lambda^2 r^2 + \alpha'} \; \qquad
 \chi_1 = \psi_1^2 = 1 \:, \qquad
 \chi_2 = \psi_2^2 = 0 \:.
\end{equation}
In addition, the $h$-equation~\eqref{3s_h_int} is explicitly solved by (or equivalently, one may take the limit $\rho\to\infty$ of eq.~\eqref{3s_h_sol}) 
\begin{equation}
 h(r) = a^{-2} + \frac{6 Q_e}{\alpha'^2} \left[ \frac{\alpha'}{r^4} - \frac{8 \lambda^2}{r^2} + \frac{32 \lambda^4}{\alpha'} \log \left(1 + \frac{\alpha'}{4 \lambda^2 r^2} \right) \right] \; .
\end{equation}
The full 10-dimensional solution can be obtained straightforwardly by plugging the two expressions above into~\eqref{3s_10dsol}. The limiting solutions for $r\to 0,\infty$ coincide with the ones in Section~\ref{sec:3s_sol} given by~\eqref{3s_sol_r_infty}-\eqref{3s_sol_r_0}.

\subsection{Topology and wrapped cycles}\label{ssec:Top+cycles}
 In the preceding four subsections we found solutions asymptotic to AdS$_3\times X^ k\times \mathbb T^{7-k}$, with
metric in Poincar\'e coordinates
 \begin{equation}
   g = \frac{s^2}{Q_e}\big(-\dd t^2 +\dd x^2\big) + Q_m \Big(\frac {\dd s^2}{s^2} + g^k\Big) + g_{\mathbb T^{7-k}},
 \end{equation} 
 up to an irrelevant coefficient in front of $\frac {\dd s^2}{s^2}$. The coordinates $(t,x,s)$ cover only a patch of
AdS$_3$, in particular the coordinate $s$ is allowed to take negative values, but $(t,x)$ are not good coordinates
around $s=0$ \cite{GibHorTown94}. The solution we have presented is valid only in the region $s>0$, since we have found
the values of the supergravity fields in this region only. Contrary to the situation for a single NS5-brane it is now
possible to continue the metric continuously beyond $s=0$, and one may wonder whether the other supergravity fields extend
as well. Clearly, this is the case for the constant dilaton, and also for the 3-form $H$, since its AdS$_3$-component is
proportional to the volume form. The gauge fields for $\psi\neq 0$ have the form
\begin{equation}
 \nabla(\psi)= \nabla^ P + \frac {\rho^2}{\rho^2 +r^2} e^aI_a,
\end{equation} 
 where $\nabla^P,e^a$ and $I_a$ are globally well-defined, and the coordinate $r$ is related to $s$ as $s = r^\vartheta$ for some
positive rational number $\vartheta$, in the region $r,s>0$. For negative values of $r$ and $s$ we can set $s=-(-r)^\vartheta$. It follows that $\nabla(\psi)$ extends continuously to negative values of $r$, and if the coordinate $r$ is smooth around $r=0$ (like $s$ is on AdS$_3$) then $\nabla(\psi)$ is smooth as well.
As we have argued in Section \ref{sec:gaugesolbrane} the limiting connection $\lim_{\psi\rightarrow 0}\nabla(\psi)$
develops a singularity at $r=0$.
But here the interpretation of the singularity is more obvious than it was for a single NS5-brane, since the metric is
continuous, and the locus $\{r=0\}$ is a 9-dimensional subspace of AdS$_3\times X^ k\times \mathbb
T^{7-k}$. It forms a horizon, where the vector field $\frac \partial{\partial r}$ becomes light-like. The brane world-volumes are located within the horizon, due to the fact that the fundamental string and NS5-branes are extremal branes \cite{HorStrom91}, i.e. their masses and charges satisfy a BPS bound.

\begin{figure}
\centerline{
        \begin{overpic}[scale=0.5]%
            {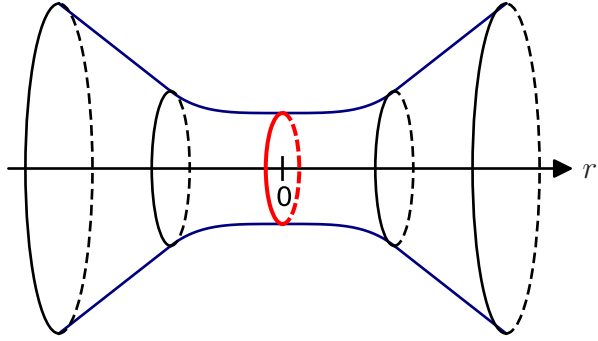} \put(225,65){\large $r$}
      \end{overpic}}
\caption{Schematic view of an (NS1+)NS5-brane geometry. The circles represent copies of $X^k$, and the brane
world-volume is localized at $r=0$. Far away from the brane the geometry looks like a cone over $X^k$, close to the brane
it is a cylinder. If only 5-branes are present, then the full space-time is a direct product of the plotted geometry
with $\mathbb R^{1,1}\times \mathbb T^{7-k}$ and the brane is at infinite geodesic distance from the finite $r$ regions.
If there are also fundamental strings inside the 5-brane then close to the brane $r$ becomes a coordinate of AdS$_3$,
the cylindrical region is asymptotic to AdS$_3\times X^k\times \mathbb T^{7-k}$ and the 9-dimensional surface $\{r=0\}$ becomes
light-like, i.e. its metric is degenerate with signature $(8,0)$. Time-like geodesics can now cross the brane in finite proper time. }\label{Fig:branecylinder}
\end{figure}

We have argued that, except for the gauge field, all fields extend at least continuously to the region $r\leq 0$, with identical solutions for $r<0$ and $r>0$, since the fields depend only on $r^2$. This is illustrated in Figure
\ref{Fig:branecylinder}. Note that a similar extension is possible when there is only an NS5-brane without strings. In this case metric and dilaton are singular at the brane location $r=0$, and the two regions $r>0$ and $r<0$ are causally disconnected. The singularities are cancelled by the addition of fundamental strings, and then time-like geodesics connect the two regions.

Tian's theorem tells us that the singular support of the limiting connection $\lim_{\psi\rightarrow 0}\nabla(\psi)$ is
a calibrated codimension four subspace (or rather a current) of our hypersurface $\{s=0\}$ \cite{Tian00,DS09}, which we
will interpret as the world-volume of the 5-brane. The calibration form for $s>0$ is given by  \cite{HN11}
 \begin{equation}
    * Q  = e^{4f}* \Big[\sfrac {\dd s}s \wedge P  + \sfrac 14 \dd P\Big].
 \end{equation} 
 Upon restriction to a submanifold $\{s=\mbox{const}\}$ this becomes $*Q\big|_{\{s=\text{const}\} }\propto *_kP$. Hence,
we expect that the world-volume of the brane is a formal sum of products of $\mathbb R^{2} \times \mathbb T^{7-k}$
with calibrated cycles of dimension $k-3$ in $X^k$, for the calibration form $*P$, localized at $s=0$. The induced
metric on the 6-dimensional world-volume is degenerate with signature $(5,0)$.
In our construction, we have not singled out any submanifolds of $X$,
and the only reasonable expectation is that all calibrated cycles get wrapped at once, so that we obtain some sort of
smeared brane. In the simplest case $X=S^3$ the calibration form is the constant function one, every point of $S^3$ is
calibrated, and the brane is smeared evenly over $S^3$. 

 For nearly K\"ahler manifolds the calibrated submanifolds are special Lagrangian, for nearly parallel $G_2$ ones they
are co-associative. In the case of Sasaki-Einstein manifolds $X$ the calibrated submanifolds of $X$ we are interested in
are complex submanifolds of the cone, when we embed $X$ into the cone as $\{1\}\times X$. It would be desirable to
construct explicitly a smooth extension of our family of instantons $\nabla(\psi)$ beyond $r=0$, which would enable us
to determine the singular support of the limiting connection with $\psi=0$ and see whether it can indeed be identified
with a union of all elementary calibrated submanifolds of $X$ localized at $r=0$.

 When we take the limit $\psi\rightarrow 0$ the gauge bundle degenerates, and has to be treated as a sheaf rather than
a bundle \cite{FrieMoWi97,Sharpe97,Tian00}. The sheaf is locally free (a vector bundle) away from the brane, but not
along the codimension four world-volume. Similarly to D-branes in type II string theory \cite{Aspinwall04} the
NS5-branes are not simply submanifolds, but come equipped with a sheaf as some extra structure.

Consider again a single NS5-brane. The metric is asymptotically cylindrical, i.e. of the form 
 \begin{equation}
   g = -\dd t^2 +\dd x^2 + \frac{\alpha'}4\Big(\frac {\dd r^2}{r^2} +g^k \Big) + g_{\mathbb T^{7-k}}
 \end{equation} 
 in the near-horizon limit.
 Usually the singular locus $\{r=0\}$ is considered as the codimension $(k+1)$-submanifold $\mathbb R^{1,1}\times \mathbb T^{7-k}$ in the full space-time, which could accommodate $p$-branes with $p\leq 8-k$ \cite{Strom90,HS90,GN95,Loginov08}. Topologically, this corresponds to the partial one point compactification of the cylinder $\mathbb R_{>0}\times X$ that gives rise to the cone. It is also possible to partially compactify the cylinder through the addition of another copy of $X$. Then we find a boundary component $\mathbb R^{1,1}\times X^k\times \mathbb T^{7-k}$ at $r=0$, which admits higher-dimensional brane world-volumes. This is indeed what we get when fundamental strings enter; since the metric and dilaton become non-singular then, there is no longer any ambiguity in the interpretation of the subspace $\{r=0\}$, which is a 9-dimensional submanifold of AdS$_3\times X^k\times \mathbb T^{7-k}$. Thus, the topology of a single NS5-brane background must be viewed as
 \begin{equation}
   \mathbb R^{1,1} \times\mathbb T^{7-k} \times \mathbb R_{\geq 0}\times X^k,
 \end{equation} 
  with a 9-dimensional boundary, rather than
 \begin{equation}
   \mathbb R^{1,1} \times\mathbb T^{7-k} \times c(X^k).
 \end{equation} 
  As mentioned before, negative values for $r$ are possible as well, but the two regions $r<0$ and $r>0$ remain causally disconnected. 

\subsection{Fundamental strings with gauge 5-branes}\label{sec:gauge5brane}
For completeness and in order to connect to the literature~\cite{Strom90,HS90,GN95,Loginov08,HN11}, we will now discuss another choice for the connections $\tilde\nabla$ and $\nabla^A$, namely $\psi_1 = 1$ and $\psi_2 >0 $, instead of $\psi_2=0$ ($\psi_1=\chi_1=1$ instead of $\psi_2=\chi_2=0$ in the Sasaki-Einstein and 3-Sasakian cases). This corresponds to fixing $\tilde\nabla$ to be the Levi-Civita connection of the cone $c(X^k)$, and gives rise to a single gauge 5-brane plus again an arbitrary number of strings. The limiting behaviour of the solutions as $r\to 0$ and $r\to \infty$ is the same as for fundamental strings only (up to a rescaling of coordinates and $\phi_0$), and in particular there is no AdS$_3$ region.

\paragraph{Nearly parallel $G_2$.}
We take the general solution from Section~\ref{sec:g2_sol} and choose $\psi_1 = 1$, $\psi_2 = \rho^2/(\rho^2 + r^2)$ with integration constant $\rho\in\mathbb{R}$. The remaining fields then read
\begin{equation}
\begin{aligned}
 e^{2f} &=\lambda^2 r^2+\frac {\alpha'}4\, \left(1 - \frac{\rho^4}{(\rho^2+ r^2)^2} \right) \; , \\
 h(r) &= a^{-2} + 6Q_e \int_{\log(r)}^{\infty} \exp\Big(-\sfrac 43 f (\theta)- \sfrac {14}3 \theta \Big) \dd\theta \; .
\end{aligned}
\end{equation} 
 As mentioned above, if $Q_e\neq 0$ the limiting behaviour as $r\to0,\infty$ is that of the fundamental string, given by \eqref{F1r=0} and \eqref{F1r=infty}, except that the gauge field $\nabla^A$ in the limit $r\to\infty$ does not coincide with $\nabla^c$. Since $\alpha'/r^2 \to0$ in this limit, $\alpha'$-corrections can be ignored and the field equations are satisfied. The gauge solitonic brane solution without strings ($Q_e=0$) for $X=S^7$ was found in \cite{HS90}. 

\paragraph{Nearly K\"ahler.}
To accommodate the choice $\psi_1 = 1$, the solution of Section~\ref{sec:nk_sol} is altered such that
\begin{equation}
\begin{aligned}
 e^{2f} &= \lambda^2 r^2 + \frac {\alpha'}4 \left(1 - \frac{\rho^4}{(\rho^2+ r^2)^2} \right) \; , \\
 h(r) &= a^{-2} + 5 Q_e \int_{\log(r)}^\infty \exp\Big(-f(\theta)-4\theta\Big) \dd\theta \; .
\end{aligned}
\end{equation} 
 The gauge solitonic brane without fundamental strings ($Q_e=0$) for $X=S^6$ was found in \cite{GN95}.

\paragraph{Sasaki-Einstein.}
For $X^{2n+1}$ Sasaki-Einstein, we need to distinguish the cases $n=1$ and $n=2,3$. For $n=1$ we set $\ell=0$, $\psi_1 = \chi_1 = 1$, $\psi \equiv \psi_2 = \chi_2$ and $Q_m=0$ (the NS5-brane is turned off in this
section), to obtain
\begin{equation}
 e^{2f} = \lambda^2 r^2 + \frac {\alpha'}4 \left(1 - \psi^2 \right) \; , \qquad \psi = \frac{\rho^2}{\rho^2+ r^2} \; , \qquad h = a^{-2} + \frac{Q_e}{r^2} \; ,
\end{equation}
and the full 10-dimensional solution~\eqref{se_10dsol} becomes
\begin{equation}
\begin{aligned}
 g &= \frac{r^2}{r^2/a^2 +Q_e}(-\dd t^2+\dd x^2) +\Big(\lambda^2 + \frac{\alpha'}4 \frac{1-\psi^2}{r^2} 
\Big) \big(\dd r^2 +r^2 g_{S^3}\big) + g_{\mathbb{T}^4},\\
 H &= \frac{2Q_er}{(r^2/a ^ 2+Q_e) ^2} \dd r \wedge \dd t\wedge \dd x + \frac{\alpha'}2 \left(1 + 2\psi^3 - 3\psi^2 \right) \mbox{Vol}_{S^3}, \\
 e^{2(\phi-\phi_0)} &= \frac {1}{r^2/a^2 + Q_e} \Big(\lambda^2 r^2+ \frac{\alpha'}4 \left(1 - \psi^2 \right) \Big).
\end{aligned}
\end{equation}
The limiting cases are again the same as for the fundamental string alone, except when $Q_e=0$, which leads to Strominger's gauge 5-brane \cite{Strom90}.

For $n=2,3$, the full solution can only be found numerically. We will refrain from numerically solving the full
eqs.~\eqref{SE_PDEsystem} and merely mention that the limiting solutions of the fundamental strings remain valid.

\paragraph{3-Sasakian.}
We take the general results from Section~\ref{sec:3s_sol} and specify
\begin{equation}
 \chi_1 = \psi_1^2 = 1 \; ,\qquad
 \chi_2 = \psi_2^2 = \frac{\rho^2}{\rho^2+r^2} \; .
\end{equation}
The full solution is then of the form as stated in~\eqref{3s_10dsol} with
\begin{align}
 e^{2f} &= \lambda^2 r^2 + \frac{\alpha^\prime}{8}(1-\chi_2^2) \; , \\
 e^{2\ell} &= 2\,\frac{4\lambda^2 r^2  + \alpha^\prime (1-\chi_2)}{8\lambda^2 r^2 + \alpha^\prime (1-\chi_2^2)} \; , \\
 h (r) &= a^{-2} + 6 Q_e \int_{\log(r)}^\infty \exp\Big(-2f(\theta)-4\theta-2\ell(\theta) \Big) \dd\theta \:.
\end{align}
 If fundamental strings are present, then they determine the limiting behaviour, otherwise we obtain a gauge solitonic brane based on an Sp(2)-instanton on the hyperk\"ahler cone $c(X^7)$. For the case $X^7=S^7$ this was first obtained in \cite{Loginov08}.

\section{Isometries and holography}\label{sec:Isoms}
 To a supergravity vacuum on a Lorentzian manifold $(M,g)$ one can associate its isometry super Lie algebra $\mathfrak
{isom}=\mathfrak b\oplus \mathfrak f$, whose bosonic part $\mathfrak b$ consists of the Killing vector fields, whereas
the fermionic part $\mathfrak f$ is spanned by the Killing spinors. It plays an important role in the AdS/CFT
correspondence, since the near horizon isometries are expected to coincide with the supersymmetry algebra of the dual
conformal field theory. The pairing that maps two Killing spinors $\psi,\epsilon$ to a Killing vector is the usual
spinor bilinear
 \begin{equation}
   \langle \epsilon, \gamma^0\gamma^\mu \psi\rangle \partial_\mu,
 \end{equation} 
 whereas the action of a Killing vector $V$ on a spinor $\epsilon$ is given by the Lie derivative \cite{FoF99}
 \begin{equation}\label{vecspinactiondefi}
   \mathcal L_V \epsilon := \nabla_V \epsilon + \sfrac 14(\nabla V^\flat)\cdot \epsilon.
 \end{equation} 
  Here $\flat$ denotes the musical isomorphism, identifying vectors with 1-forms via the metric, with inverse mapping
denoted by $\sharp$. The covariant derivative $\nabla V^\flat $ for $V$ a Killing vector is a 2-form, which acts
naturally on spinors.
 In the present situation the Killing spinor $\epsilon$ is parallel with respect to a metric connection with torsion
proportional to the 3-form $H$, and we can rewrite the action \eqref{vecspinactiondefi} in terms of the 1-form
$\theta= V^\flat$ as
 \begin{equation}\label{vecspinaction2}
   \mathcal L_{\theta^\sharp} \epsilon = \frac 14 \big( \nabla \theta + \theta \lrcorner H\big) \cdot \epsilon =\frac 14
(\nabla^+ \theta)\cdot \epsilon,
 \end{equation} 
 where $\nabla^+$ is the metric compatible connection with torsion equal to minus the torsion of $\nabla^-$.
 Since 2-forms on a pseudo-Riemannian manifold of signature $(p,q)$ can be identified with the Lie algebra
$\mathfrak{so}(p,q)$, we find that the isometries acting non-trivially on Killing spinors form a subgroup of Spin(9,1).
As an example consider a connected simply connected simple Lie group $G$ equipped with its bi-invariant metric. The group $G$
carries a bi-invariant 3-form $H$, defined by
 \begin{equation}
   H(X,Y,Z) = -\langle X,[Y,Z]\rangle
 \end{equation}
 for $X,Y,Z$ elements of the Lie algebra. The holonomy groups of both $\nabla^\pm$ are trivial, left-invariant vector
fields and left-invariant spinors on $G$ are parallel with respect to $\nabla^-$, and right-invariant vector fields and
spinors are parallel with respect to $\nabla^+$ \cite{KoNoII}. Hence, the gravitino equation is solved by all
left-invariant spinors on $G$. It follows from \eqref{vecspinaction2} that right-invariant vector fields act trivially
on the Killing spinors, so if we neglect any purely bosonic part of the isometry group then the remaining isometries are
generated by all left-invariant vector fields and left-invariant spinors.

Note that a Lie group with its bi-invariant metric and 3-form does not by itself solve the supergravity
equations, since the dilatino equation does not hold, but it can appear as a factor in the solution. The near horizon
AdS$_3$ solutions are exactly of this type,
with $G =SL(2,\mathbb R) = \mbox{AdS}_3$. Thus, the non-trivial part of the isometry algebra of our near horizon
backgrounds contains one $\mathfrak {sl}(2,\mathbb R) = \mathfrak{so}(2,1)$ component, similarly to the proposed
maximally supersymmetric AdS$_3\times S^k\times \mathbb T^{7-k}$ solutions of \cite{LSS07,KLS07}. In the following we will determine the super isometries of the near horizon solutions, neglecting bosonic isometries that act trivially on the Killing spinors, like the right-invariant Killing vectors on
AdS$_3$.\vspace{\baselineskip}

\paragraph{Nearly K\"ahler and nearly parallel $G_2$.} In this case the isometry algebra is simply $\mathfrak{so}(2,1)$,
and the spinors are in the 2-dimensional representation. The resulting super Lie algebra is $\mathfrak{osp}(1|2)$
\cite{Frappat96}.

\paragraph{3D Sasaki-Einstein.}
The only simply connected 3-dimensional Sasaki-Ein\-stein manifold is the 3-sphere $S^3
=\mbox{SU}(2)$. Similarly to the AdS$_3$ component, its bosonic isometries are SU(2$)\times \mbox{SU}(2)$ but only one
SU(2) acts non-trivially on spinors. There are sixteen $\nabla^-$-parallel spinors, but only eight of them solve the dilatino equation.
The eight Killing spinors transform in the $(\mathbf{2},\mathbf{2}) \oplus
(\mathbf{2},\overline{\mathbf{2}})$ representation of SL($2,\mathbb R)\times\mbox{SU}(2)$, and the resulting super Lie
algebra is $\mathfrak{psu}(1,1|2)$. This result also follows from the analogous
solution of type IIB supergravity,
describing the horizon of a D1+D5-brane system, with isometries $\mathfrak{psu}(1,1|2)\oplus
\mathfrak{psu}(1,1|2)$ \cite{Mald07,Claus98}, and whose world-sheet conformal field theory is the product of an
SL($2,\mathbb
R)$- with an SU(2)-WZW model \cite{Tseytlin96}. Note also that \cite{LSS07} proposed the isometry algebra
$D(2,1;\alpha)\oplus D(2,1;\alpha)$ for the expected heterotic AdS$_3\times S^3 \times \mathbb T^4$ solution with
maximal supersymmetry, similarly to the M-theory case \cite{Boonstra98}.

\paragraph{5D and 7D Sasaki-Einstein.} The bosonic isometry algebra is $\mathfrak{so}(2,1)\oplus \mathfrak u(1)$.
Sasaki-Einstein manifolds come equipped with a 1-form $\eta$ and a 2-form $\omega$ such that $\nabla\eta = 2\omega$, and
the 3-form $H$ is given by $2\eta\wedge \omega$. The $\mathfrak u(1)$ isometries are generated by the vector field dual
to $\eta$, and we have 
 \begin{equation}
   \mathcal L_{\eta^\sharp} \epsilon =  \omega \cdot \epsilon. 
 \end{equation} 
 The resulting super Lie algebra is $\mathfrak{osp}(2|2)$.

\paragraph{3-Sasakian.} The Killing vectors form an $\mathfrak{so}(3)$-algebra. A 3-Sasakian manifold has three 1-forms
$\eta^\alpha$ and three 2-forms $\omega^\alpha$, $\alpha=1,2,3$, such that 
 \begin{equation}
   \nabla \eta^\alpha = 2\omega^\alpha + {\varepsilon^\alpha}_{\beta\gamma} \eta^\beta\wedge \eta^\gamma.
 \end{equation} 
 The 3-form is given by $H = 2\eta^\alpha\wedge \omega^\alpha + 2 \eta^{123}$, and the $\mathfrak{so}(3)$-Killing
vectors are generated by the $(\eta^\alpha)^\sharp$. We conclude that
 \begin{equation}
   \mathcal L_{(\eta^\alpha)^\sharp} \epsilon = \big(\omega^\alpha + \sfrac 12{\varepsilon^\alpha}_{\beta\gamma}
\eta^\beta \wedge \eta^{\gamma}\big) \cdot \epsilon.
 \end{equation}  
 It follows from the results of \cite{HN11} that the Killing spinors transform in the $\mathbf{3}$-representation of
$\mathfrak{so}(3)$, and the full isometry algebra is $\mathfrak{osp}(3|2)$.

 \renewcommand{\arraystretch}{1.2}
  \begin{table}[H]\centering
 \begin{tabular}{c|c|c|c|c}\toprule
   $\dim X$ & $X$ & $\mathfrak b$ & $\mathfrak f$ &  $\mathfrak{isom}$ \\ \hline 
 7 &  nearly parallel $G_2$&  $\mathfrak{so}(2,1) $ & $(\mathbf{2})$& $\mathfrak{osp}(1|2)$  \\
5, 7 & Sasaki-Einstein & $\mathfrak{so}(2,1)\oplus \mathfrak {u}(1)$ & $(\mathbf{2})\oplus (\overline{\mathbf{2}})$
&$\mathfrak{osp}(2|2)$ \\
  7 & 3-Sasakian &  $\mathfrak{so}(2,1)\oplus \mathfrak {sp}(1)$ & $(\mathbf{2},\mathbf{3})$ &  $\mathfrak{osp}(3|2)$  \\ 
  6 & nearly K\"ahler & $\mathfrak{so}(2,1)$  & $(\mathbf{2})$&  $\mathfrak{osp}(1|2)$ \\
3 & Sasaki-Einstein & $\mathfrak{so}(2,1) \oplus \mathfrak{su}(2)$ & $(\mathbf{2},\mathbf{2})\oplus (\mathbf{2},\overline{\mathbf{2}})$ &
$\mathfrak{psu}(1,1|2)$
\\ \bottomrule
\end{tabular}
 \caption{Super isometry algebras $\mathfrak{isom}=\mathfrak b\oplus \mathfrak f$ of the near horizon solutions.}\label{tab:sisoms}
\end{table}
According to Brown and Henneaux, the isometry algebra of an asymptotic AdS$_3$ geometry admits an affine extension
containing two Virasoro algebras \cite{BH86}, in the supergravity case a possibly non-linear superconformal algebra
\cite{HMS99}. The affine extension comes from bulk diffeomorphism that do not vanish rapidly at the
conformal boundary. The classification of these superconformal algebras can be found in \cite{Fradkin92,Bowcock92} (or
see \cite{KLS07}); the ones relevant to our cases are 
 \begin{equation}
   \widehat{\mathfrak{osp}}(N|2) \und \widehat{\mathfrak{psu}}(1,1|2),
 \end{equation} 
 for $N=1,2,3.$ These are classical linear superconformal algebras, except for $\widehat{\mathfrak{osp}}(3|2)$ which
contains a non-linear term in the commutation relations. They should be compared to the following superconformal algebras, which have been proposed as isometries of maximally supersymmetric AdS$_3$-backgrounds in \cite{DabhMurthy07,LSS07,KLS07}: 
 \renewcommand{\arraystretch}{1.2}
  \begin{table}[H]\centering
 \begin{tabular}{cc}\toprule
    geometry & $\mathfrak{isom}$ \\ \hline 
  AdS$_3\times S^7$ &  $\mathfrak{osp}(8|2)$ \\
AdS$_3\times S^6$ &  $\mathfrak{f}(4)$  \\
  AdS$_3\times S^5$ &  $\mathfrak{su}(1,1|4)$ \\ 
  AdS$_3\times S^4$ &  $\mathfrak{osp}(4^*|4)$  \\
AdS$_3\times S^3$ &  $D(2,1;\alpha)\oplus D(2,1;\alpha)$\\
 AdS$_3\times S^2$ &  $ \mathfrak{osp}(4^*|4)$\\ \bottomrule
\end{tabular}
 \caption{Proposed super isometry algebras for hypothetical maximally supersymmetric solutions of
heterotic supergravity.}\label{tab:propsisoms}
\end{table}
 All of the superalgebras in Table \ref{tab:propsisoms} possess superconformal extensions as well, with $\widehat
D(2,1;\alpha)$ being the only linear super Lie algebra, but $\widehat{\mathfrak{osp}}(4^*|4)$ and
$\widehat{\mathfrak{su}}(1,1|4)$ cannot have unitary highest weight representations \cite{Knizhnik86,Bowcock92,KLS07}, a
problem that does not seem to occur for the algebras relevant to our backgrounds. The superconformal algebras with a simple compact bosonic subalgebra possess a discrete level $k$, which was shown to be related to the number of strings $N$ in \cite{KLS07}. In the case of $\widehat{\mathfrak{osp}}(1|2)$ with its bosonic subalgebra $\widehat{\mathfrak{sl}}(2,\mathbb R)$ the level may assume continuous values, however.

Since the supergravity backgrounds asymptote to AdS$_3$ we expect them to possess holographically dual 2-dimensional conformal field theories, whose symmetries should coincide with the super isometries of the near horizon geometry. There is an obvious candidate for the CFT side with the right symmetries, namely a world-sheet sigma model with target space the near horizon background. Isometries of the target space give rise to symmetries of the sigma model (e.g. \cite{HullSpence89} for the classical sigma model), hence the bosonic symmetries match. Furthermore, the occurrence of a `heterotic' isometry algebra, i.e. one with left-moving but no right-moving supersymmetry, is a strong indication of a heterotic world-sheet theory. 
 
The simplest example is the sigma model with target AdS$_3\times S^3$, which is a Wess-Zumino-Witten (WZW) model on
SL($2,\mathbb R)\times \mbox{SU}(2)$, with vanishing $\alpha'$-corrections.
The IIB WZW model was shown explicitly to admit a $\widehat {\mathfrak{psu}}(1,1|2) \oplus \widehat {\mathfrak{psu}}(1,1|2)$ symmetry algebra \cite{GivKutSei98,MaldOoguri}, so that we can indeed expect to find an $\widehat {\mathfrak{sl}}(2,\mathbb R)\oplus \widehat{\mathfrak{su}}(2) \oplus \widehat {\mathfrak{psu}}(1,1|2)$ algebra in the heterotic setting.
Proposals for the heterotic world-sheet CFT on maximally supersymmetric AdS$_3$ backgrounds have been made in
\cite{DabhMurthy07,LSS07,Johnson07}.

\section{Conclusion}

In this paper we have presented order $\alpha'$ solutions of heterotic supergravity based on space-time patches of the form
$\mathbb{R}^{1,1} \times \mathbb{R} \times X^k \times \mathbb{T}^{7-k}$, with $X^k$ being one of the four types of
geometric Killing spinor manifolds in dimensions $k =3,5,6,7$. The solutions describe the intersection of a fundamental
string with an NS5-brane and an optional gauge anti-5-brane, and generalize the previously known case of NS1+NS5-branes
on flat Minkowski space, where $X=S^3$. We have found the complete analytical solution for nearly parallel $G_2$ and
nearly K\"ahler manifolds $X$ in the case of
absent gauge anti-5-brane, and the complete solution for 3-Sasakian manifolds. The NS5-brane wraps calibrated cycles of
dimension $k-3$ inside $X^k$, which is a common property of BPS branes \cite{Gaunt03}, and it would be interesting to
determine the cycles in some special cases.
Furthermore, on the brane world-volume the gauge bundle associated to the NS5-brane degenerates to a sheaf, which comes equipped with a
`point-like instanton' or its higher dimensional generalization. Contrary to the case $X=S^3$ the singular instanton is
supported not only on the world-volume, but coincides with the pull-back of the canonical connection on $X$ away
from the brane.

 Motivated by the separate supergravity solutions for fundamental strings and NS5-branes, we have made the following
ansatz for the 10-dimensional metric $g$ and connection $\nabla^-$ (in the
string frame; the choice of frame is relevant for the structure in which
the functions $h$ and $f$ appear in the metric.):
\begin{equation}
\begin{aligned}
 g  &= h^{-1}(\tau) \big(-\dd t^2 +\dd x^2 \big) + e^{2f(\tau)}\big(\dd \tau^2 + g^k(\tau)\big) + g_{\mathbb{T}^{7-k}}
\; , \\
 \nabla^- &= \nabla^ P + s_{ab}(\tau) e^a I_b + e^{-f}\partial_\tau(h^{-1/2}) \big(\dd t-\dd x\big) I_8 - \sfrac
14\dd(\log h) Z \; .
\end{aligned}
\end{equation}
Here $\nabla^P$ is the canonical connection on $X^k$, with totally antisymmetric torsion and reduced holonomy. $I_8$ is a generator of the
$\mathbb R^8$ subalgebra of $\mathfrak{spin}(9,1)$, $Z$ a generator of $\mathfrak{so}(1,1)$, the $e^a$ are a basis of
1-forms on $X^k$, and the $I_b$ generate the orthogonal complement of the structure group of $X^k$ inside the holonomy
group of the cone $c(X^k)$. The coefficient matrix $s_{ab}(\tau)$ is in fact diagonal, and for a nearly
K\"ahler or nearly parallel $G_2$-manifold $X^k$ it is proportional to the unit matrix. In this case the internal metric
$g^k$ is $\tau$-independent, whereas it has some explicit $\tau$-dependence if $X^k$ comes equipped with a Sasakian
structure. 
The gauge connection $\nabla^A$ is identified with the pull-back of the canonical connection $\nabla^P$ of $X$ away from the horizon at $r=0$ ($r=e^\tau$) where it becomes singular, and it gives rise to the NS5-brane. The additional gauge anti-5-brane is obtained when the connection $\tilde \nabla$ interpolates between $\nabla^P$ as $r\rightarrow \infty$ and $\nabla^c$ for $r\rightarrow 0$, and it is absent when $\tilde \nabla=\nabla^c$ globally.

Solving the heterotic BPS equations~\eqref{BPSeqtns}, the Bianchi identity~\eqref{Bianchi}
and the time-like components of the field equations~\eqref{eom} determines the remaining degrees of freedom in $g$ and
$\nabla^-$, as well as fixing the other 10-dimensional bosonic fields $H$ and $\phi$. The most important deviation from the separate solutions for fundamental strings and 5-branes is that the function $h$ is no longer harmonic for an NS1+NS5-brane system. On the other hand, the field components coming from the 5-branes remain unchanged.
 The solutions interpolate between
\begin{equation}
 \mbox{AdS}_3\times X^k \times\mathbb{T}^{7-k} \qquad \rightarrow \qquad \mathbb{R}^{1,1} \times c(X^k) \times\mathbb{T}^{7-k} \; ,
\end{equation}
whereby the 10-dimensional fields take the explicit form 
\begin{equation}\label{eq:summary_sol_r0}
\begin{aligned}
 g  &= s^2 (-\dd t^2 +\dd x^2 ) + \varepsilon_1 \alpha' \frac{\dd s^2}{s^2} + \varepsilon_2 \frac{\alpha'}{4} g^k +
g_{\mathbb{T}^{7-k}} \qquad (s\propto r^{\varepsilon_5})\: ,\\
 H&= \dd (s^2) \wedge \dd t \wedge \dd x + \varepsilon_3 \frac{\alpha'}{2} P, \qquad\quad
 e^{2(\phi-\phi_0)} = \frac{\varepsilon_4}{Q_e} \left(\frac{\alpha'}{4}\right)^{(k-1)/2}  \; ,
\end{aligned}
\end{equation}
as $r\to 0$. The AdS$_3$ radius is determined by $\varepsilon_1$ as $R_{\text{AdS}_3}^2 = \varepsilon_1 \alpha'$. For $r\to \infty$ we have
\begin{equation}\label{eq:summary_sol_rinfty}
\begin{aligned}
 g &= a^2(-\dd t^2+\dd x^2)  +\lambda^2 (\dd r^2 +r^2 g^k) + g_{\mathbb{T}^{7-k}}\\
 H&= 0, \qquad e^{2(\phi-\phi_0)} = a^2\lambda^{\varepsilon_5}.
\end{aligned}
\end{equation}
This succinctly summarizes the main results of Section~\ref{sec:oursol}. In their limiting behaviour, the solutions for the four types of geometric Killing spinor manifolds share the same structure differing only by some numerical coefficients $\varepsilon_1$, $\ldots$, $\varepsilon_5$. The latter are collected in Table~\ref{tab:consts}.
 \renewcommand{\arraystretch}{1.2}
  \begin{table}[H]\centering
 \begin{tabular}{cccccc}  \toprule
 $X^k$ & $\varepsilon_1$ & $\varepsilon_2$ & $\varepsilon_3$ & $\varepsilon_4$ & $\varepsilon_5$ \\ \hline
 nearly parallel $G_2$ &	$9/196$ &	$1$ &			$1/3$ &			$7/9$ &			$14/3$		\\
 nearly K\"ahler &		$1/16$ &	$1$ &			$1/2$ &			$4/5$ &			$4$		\\
 Sasaki-Einstein &		$2/(k^2-1)$ &	$\frac{k+1}{2(k-1)}$ &	$\frac{k+1}{2(k-1)}$ &	$\frac{k+1}{2(k-1)}$ &	$(k+1)/2$	\\
 3-Sasakian &			$1/32$ &	$1/2$ &			$3/2$ &			$1/3$ &			$4$		\\
 \bottomrule
\end{tabular}
\caption{Numerical coefficients for the limiting solutions~\eqref{eq:summary_sol_r0}-\eqref{eq:summary_sol_rinfty} for the four types of geometric Killing spinor manifolds.}\label{tab:consts}
\end{table}
Perturbative string theory can be thought of as a double expansion in the string coupling $g_s = e^\phi$ and the Regge
slope $\alpha'$. For a large number of fundamental strings our solutions have a small string coupling, so that we can
trust the first expansion. The explicit $\alpha'$-dependence of the solutions, however, gives rise to large correction
terms at higher order in $\alpha'$, so that there is a problem with this expansion. This may be resolved by considering
multiple 5-branes instead of a single one. For instance, equation \eqref{S3nhlimit} shows that in the near
horizon limit for $X=S^3$ both the AdS$_3$ and $S^3$ radii are proportional to $Q_m^{1/2}$, whereas the dilaton
assumes the form $e^{2(\phi-\phi_0)} =  Q_m/Q_e$. Here $ Q_m$ is the magnetic NS5-brane charge, proportional to the number of branes, and $Q_e$ is the electric charge of the fundamental strings. Hence, to keep the string coupling
small and the volumes large we need to have a large number of 5-branes, while the ratio of 5-branes to strings must
remain small. Gauge multi-5-branes on Minkowski space from $G_2$- and Spin(7)-instantons have been constructed
in \cite{Loginov05}. In the small instanton limit they should give rise to multiple NS5-branes.

For $X=S^3$ one can even construct a superposition of fundamental strings with an arbitrary number of gauge 5-branes and
NS5-branes \cite{LLOP99}, and for a particular choice of the parameters one obtains a solution that interpolates between
two AdS$_3\times S^3$ regions with different radii, which has been interpreted as a dual gravitational theory to a
renormalization group flow of a conformal field theory \cite{Gava10}. Similar solutions with $S^3$ replaced by some
arbitrary manifold with geometric Killing spinors can be expected to exist, but require the construction of a
superposition of gauge 5-branes with NS5-branes, which we have not succeeded in so far.
We have also considered the intersection of a fundamental string with a gauge 5-brane only. In this case we do not find
an asymptotic AdS$_3$ region. The transformation behaviour of NS1+NS5-brane systems under string dualities has been
discussed in \cite{PapRusTseyt00}; generalizing this method to our solutions should yield new curved brane solutions of
other supergravities.

The holographic properties of the supergravity solutions have not been studied in this work, except for the super isometry algebras of the near horizon geometries, which we determined. They are of `heterotic' type, and hence give rise to the expectation that the dual CFT is a heterotic world-sheet theory. Clearly, this deserves further study.

 We have found the explicit supergravity solution describing the intersection of a fundamental string and a 5-brane on the cone over a manifold with a geometric real Killing spinor, generalizing earlier constructions on Minkowski space. It is not clear whether this is the most general situation where our construction can be applied. Consider for instance an
arbitrary non-compact Ricci-flat manifold $M$ with a codimension one submanifold $X$. If $M$ has a parallel spinor
$\epsilon$, then its restriction to $X$ satisfies a generalized geometric Killing spinor equation
 \begin{equation}\label{concl:genKillingspinor}
   \nabla_{V} \epsilon - \frac {i} 2 A(V) \cdot \epsilon=0 \qquad \forall\, V\in \Gamma(TX),
 \end{equation} 
 where $A\in \Gamma($End$(TX))$ is the Weingarten tensor of $X$ \cite{Baer_generalizedcylinder}. Like the ordinary
geometric Killing spinor equation it implies restrictions on the torsion classes of $X$. For instance, if $M$ has
holonomy Spin(7) then $X$ carries a cocalibrated $G_2$-structure, if $M$ has holonomy $G_2$ then $X$ is a half-flat
SU(3)-manifold, and if $M$ is 6-dimensional Calabi-Yau then $X$ has a so-called hypo SU(2)-structure
\cite{Hitchin01,ChiossiSal02,ContiSal05,CLSS09}. Of course, these geometric structures on $X$ are generalizations of the
nearly parallel $G_2$, nearly K\"ahler and Sasaki-Einstein geometries, which have $A =\mbox{id}$. It remains an interesting
question whether the brane solutions presented in this work can be generalized to the case where $M$ admits an arbitrary foliation by codimension one submanifolds, with a spinor satisfying \eqref{concl:genKillingspinor}, or whether this requires additional restrictions on $A$. 

\section*{Acknowledgements}
This work was partially supported by the Deutsche
Forschungsgemeinschaft, the cluster of excellence QUEST, the
Heisenberg-Landau program and Russian Foundation for Basic Research. In
particular, K.-P.G. and C.N. are grateful to the Graduiertenkolleg GRK
1463 `Analysis, Geometry and String Theory' for support.

\end{document}